\documentclass[aps,prl,twocolumn,showpacs,superscriptaddress,longbibliography]{revtex4-2}

\usepackage{amsmath}
\usepackage{graphicx}
\usepackage{lmodern}
\usepackage{amsmath}

\usepackage{color}
\usepackage{amssymb}
\usepackage{bm}
\usepackage{braket}
\usepackage{mathtools}
\usepackage{float}

\DeclareUnicodeCharacter{2215}{}
\DeclareUnicodeCharacter{03BA}{}
\DeclareUnicodeCharacter{0393}{}

\usepackage[dvipsnames]{xcolor}
\usepackage[colorlinks=true]{hyperref}
\hypersetup{
    colorlinks=true,
    linkcolor=blue,
    citecolor=blue,
    urlcolor=blue
} 

\makeatletter
\def\maketitle{
\@author@finish
\title@column\titleblock@produce
\suppressfloats[t]}
\makeatother
\begin{document}

\title{Single-band Triangular Lattice Hubbard Model with Tunable Anisotropy \\from Twisted Diamond Homobilayers}

\author{Wen Sun}
\thanks{These two authors contributed equally to this work.}
\affiliation{Institute for Advanced Study, Tsinghua University, Beijing 100084, China}

\author{Chuyi Tuo}
\thanks{These two authors contributed equally to this work.}
\affiliation{Institute for Advanced Study, Tsinghua University, Beijing 100084, China}

\author{Hong Yao}
\email{yaohong@tsinghua.edu.cn}
\affiliation{Institute for Advanced Study, Tsinghua University, Beijing 100084, China}

\begin{abstract}
The ground-state properties of the single-band triangular lattice Hubbard model with hopping anisotropy and strong interactions remain elusive so far. Here we show that twisted diamond homobilayers with band extrema at $Y$ valley can realize weakly-coupled chains with quasi-1D band structure; applying displacement field generates interchain hopping, transforming this quasi-1D system into a 2D one. The low-energy physics can be described by localized Wannier functions on the  triangular lattice with tunable hopping anisotropy, providing a promising platform for studying the anisotropic triangular lattice Hubbard model. We further employ density matrix renormalization group to study this model with interaction $U=10t$ and anisotropy $0.5\leq t'/t\leq 1.5$ at half filling, and obtain a rich ground state phase diagram, including a chiral spin liquid phase, non-magnetic phases, and a Néel antiferromagnetic phase. This work provides a first realization of displacement-field tuned anisotropy in a single-band triangular Hubbard model within moiré systems, establishing them as a promising platform to investigate intriguing correlated physics with tunable anisotropy.
\end{abstract}
\date{\today}

\maketitle
{\bf Introduction:}
Moiré superlattice systems \cite{mircroscopicofmoire,Kennes:2021aa,reviewofmoiregraphene,reviewofmoireTMD,Castellanos-Gomez:2022aa} have emerged as promising platforms for simulating strongly correlated electrons \cite{moirebandstructureprl,macdonaldmodel,TBGinsulator,TBGsc,TBGorbitalmagnet,nematicTBG2,nematicTBG1,nematicscTBG,cherninsulatorTBG,Xie:2021aa,TMDhubbard,TMDTI,hubbardTMDexperiemnt,correlatedinsulatorTMD,MottTMD,WignercrystalTMD1,chinaFQHE,QAHETMD,FQAHETMD3,FQAHETMD2,FQAHETMD1,heavyfermionTMD,FQSHETMD,SCTMD2,SCTMD1}, which offer versatile tuning knobs for precise control over band structures and interactions, such as twist angle, displacement field and filling. 
Furthermore, various experimental techniques can be applied to gain comprehensive insights into the nature of these systems. 
As a fundamental model, the isotropic single-band triangular lattice Hubbard model has been successfully experimentally realized in moiré superlattice systems \cite{hubbardTMDexperiemnt,correlatedinsulatorTMD,MottTMD,WignercrystalTMD1}, with a Mott insulating phase observed at half filling.

An important yet unresolved aspect of the single-band triangular lattice Hubbard (or Heisenberg) model is the role of hopping anisotropy, which is believed to be crucial in capturing the physics of quantum spin liquid (QSL) candidate organic crystals, such as $\kappa$-ET or Pd(dmit)$_2$ compounds \cite{Geiser:1991aa,organicmott,Etcompound,NMRET,annurevorganic,IOPrevieworganic,Yamashita:2008aa,Yamashita:2010aa,Hassan:2018aa,PhysRevX.9.041051,PhysRevLett.123.247204,Watanabe:2012aa}. Though extensive numerical studies have tried to tackle this issue \cite{chiraltriangular,PhysRevB.110.L041113,PhysRevB.106.094420,PhysRevB.91.245125,diracj1j2,chiralj1j2j3,PhysRevLett.125.157002,Peng:2021aa,dopetj,dopehubbard,PhysRevB.109.L081116,PhysRevB.102.115136,PhysRevB.96.205130,PhysRevLett.127.087201}, the underlying physics is still under debate at the current stage. And experimentally, anisotropy is typically constrained to a limited set of values, as it is determined by the underlying materials, which hinders a comprehensive understanding of its role. Considering the high degree of control available in moiré systems, a natural question that arises is, whether moiré superlattices can be engineered to realize a single-band triangular lattice Hubbard model with tunable anisotropy.

Recently, several groups have revealed moiré band structures in twisted semiconductor homobilayers with band extrema at $M$ valleys on square or triangular lattices \cite{tunablehubbardmodel,newtwistedM,twistedBC3,jingwangmvalley,macdonaldMvalley,luo2025singlebandsquarelatticehubbard}. A common feature of these moiré structures is the existence of layer exchange symmetry in the continuum model, which leads to decoupled sublattices in the low-energy description. 
The application of displacement field breaks this symmetry \cite{tunablehubbardmodel,luo2025singlebandsquarelatticehubbard}, 
effectively inducing hoppings between different sublattices.
Motivated by this mechanism, it is tempting to realize a single-band triangular lattice model with tunable hopping anisotropy controlled by the displacement field.

\begin{figure}[t]
\includegraphics[width=1.0\linewidth]{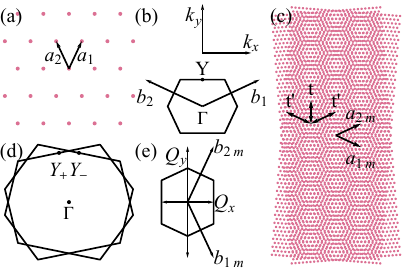}
\caption{\label{fig:moire}(a) The Bravais diamond lattice of monolayer material. (b) The first Brillouin zone of the Bravais diamond lattice. The low-energy states reside at the valley $\bm{Y}$. (c) The moiré pattern generated from twisting two Bravais diamond lattices and its primitive moiré lattice vectors. $t$ and $t'$ represents the hopping amplitudes of low-energy moiré band. (d) The Brillouin zones of top and bottom layer. (e) The moiré reciprocal lattice vectors and the wave vectors of interlayer tunneling harmonics.}
\end{figure}

In this Letter, we investigate the moiré band structure of twisted semiconductor homobilayers with band extrema at the $Y$ valley on diamond lattices. Based on a symmetry-derived continuum model, we demonstrate that the low-energy moiré band can be effectively described as weakly-coupled chains.  
Neighboring chains, which possess opposite eigenvalues under emergent layer exchange symmetry, can be coupled by the application of a displacement field. Consequently, the hopping anisotropy of electrons becomes tunable via the displacement field. The low-energy narrow band is described by localized Wannier orbitals arranged on the triangular lattice, suggesting that correlation effects in the system can be captured by a single-band triangular Hubbard model with tunable anisotropy. We employ density matrix renormalization group (DMRG) to study this model and find a chiral spin liquid (CSL) phase \cite{PhysRevLett.59.2095,PhysRevB.39.11413,PhysRevLett.99.097202,PhysRevLett.99.247203}, non-magnetic (NM) phases, and a Néel antiferromagnetic (AFM) phase that is tunable through displacement field. Our approach provides a promising route to investigate correlated physics driven by hopping anisotropy.

{\bf Model:}
We consider a basic setup where two identical layers of Bravais diamond lattices are twisted relative to each other from the AA stacking configuration by a small twist angle $\theta$, as shown in Fig.~\ref{fig:moire}(c). 
The primitive lattice vectors of the Bravais diamond lattice $\bm{a}_1$ and $\bm{a}_2$ are equal in length (Fig.~\ref{fig:moire}(a)) \footnote{We exclude angles between $\bm{a}_1$ and $\bm{a}_2$ that result in square and isotropic triangular lattices.}, and the reciprocal lattice vectors $\bm{b}_1$ and $\bm{b}_2$ defines its first Brillouin zone (Fig.~\ref{fig:moire}(b)). We aim to study the twistronics of Bravais diamond lattice semiconductor with band extrema at the valley $\bm{Y}=(\bm{b}_1+\bm{b}_2)/2$, which is invariant under both time-reversal and layer group symmetries. For simplicity, we focus on the valence band maximum (VBM) in this work (the theory of conduction band minimum can be similarly derived or related by a particle-hole transformation).
Moreover, effective spin $SU(2)$ symmetry naturally emerges at $Y$ valley due to Kramers degeneracy if both time-reversal symmetry and spatial ($C_{2z}$ or inversion) symmetry are present.
We therefore omit the spin index in the following for simplicity, as it only introduces a two-fold degeneracy. 

We first consider the kinetic energy part. 
The low-energy holes near $\bm{Y}$ with crystal momentum $\bm{Y}+\bm{p}$ exhibit the following quadratic dispersion \footnote{Layer group symmetry constrains the effective mass tensor to be diagonal along the principal axes.},
\begin{equation}\label{hp}
    h(\bm{p})=-\left(\frac{p_x^2}{2m_x}+\frac{p_y^2}{2m_y}\right)
\end{equation}

The top $(l=1)$ and bottom $(l=-1)$ layers are twisted by angle $l\frac{\theta}{2}$. We denote the state of layer $l$ with crystal momentum $\bm{p}+\bm{Y}_l$ as $|\bm{p},l\rangle$, where $\bm{Y}_l=R(l\theta/2)\bm{Y}$ (Fig.~\ref{fig:moire}(d)) is the valley base momentum rotated by $l\theta/2$. 
The kinetic energies on two layers can be written as $h(R(-l\theta/2)\bm{p}) \approx h(\bm{p})$, where we have neglected the twist angle effects on the kinetic energy at small twist angle regime. 

The interlayer tunneling matrix element between states $|\bm{p},l\rangle$ and $|\bm{p}',-l\rangle$ obeys a selection rule derived from momentum conservation,
\begin{equation} 
\bm{p}+\bm{Y}_l+\bm{G}_l=\bm{p}'+\bm{Y}_{-l}+\bm{G}'_{-l}
\label{eq:momentumconservation}
\end{equation}

In layer $l$, the electron momentum is conserved up to reciprocal lattice vectors $\bm{G}_l=R(l\theta/2)\bm{G}$, spanned by rotated basis vectors $R(l\theta/2)\bm{b}_{1}$ and $R(l\theta/2)\bm{b}_{2}$. In the low-energy theory, $|\bm{p}|$ and $|\bm{p}'|$ are smaller than the monolayer Brillouin zone, restricting $\bm{G}=\bm{G}'$ in Eq.~\ref{eq:momentumconservation}. Therefore, the interlayer momentum transfer is quantized as,
\begin{equation}
 \bm{p}-\bm{p'}=2\sin(\theta/2)(\bm{Y}+\bm{G})\times\hat{z}
 \end{equation}
Due to the small twist angle $\theta$, electrons are modulated by a long-wavelength moiré potential during tunneling. The two smallest scattering wave vectors, $\bm{Q}_{x}=2\sin(\theta/2)(\bm{b}_2+\bm{b}_1)/2\times\hat{z}$ and $\bm{Q}_{y}=2\sin(\theta/2)(\bm{b}_2-\bm{b}_1)/2\times\hat{z}$, are shown in Fig.~\ref{fig:moire}(e). Interlayer tunneling restricts electron momentum is conserved only up to $2\sin(\theta/2)\bm{G}\times\hat{z}$ in one layer. The moiré reciprocal lattice vectors $\bm{b}_{1m}=2\sin(\theta/2)\bm{b}_{1}\times\hat{z},\bm{b}_{2m}=2\sin(\theta/2)\bm{b}_{2}\times\hat{z}$ and the corresponding moiré lattice vectors $\bm{a}_{1m}=1/(2\sin(\theta/2))\bm{a}_{1}\times\hat{z},\bm{a}_{2m}=1/(2\sin(\theta/2))\bm{a}_{2}\times\hat{z}$ are shown in Fig.~\ref{fig:moire}(e) and (c), respectively.

Interlayer tunneling magnitude decreases rapidly with momentum transfer in realistic materials \cite{macdonaldmodel}. Therefore, to capture the two-dimensional moiré potential, we adopt the two-harmonics approximation, retaining only the interlayer tunneling terms with momentums $\pm\bm{Q}_{x}$ and $\pm\bm{Q}_{y}$. For convenience, we represent the Hamiltonian in real continuum space using Fourier-transformed states $|\bm{r},l\rangle\equiv\int_{\bm{p}}e^{-i\bm{p}\cdot\bm{r}}|\bm{p},l\rangle$, where $\int_{\bm{p}}\equiv\int\frac{d^2p}{(2\pi)^2}$. The interlayer tunneling  Hamiltonian is $\int d\bm{r}|\bm{r},+\rangle T(\bm{r})\langle\bm{r},-|+\text{H.c.}$, where the general form of $T(\bm{r})$ is \cite{supp},
\begin{equation}\label{tr}T(\bm{r})=2w_x\cos(\bm{Q}_x\cdot\bm{r}+\phi_x)+2w_y\cos(\bm{Q}_y\cdot\bm{r}+\phi_y)
 \end{equation}
where $T(\bm{r})$ is invariant under translations by $\bm{a}_{2m}\pm\bm{a}_{1m}$.
 
Within the two-harmonics approximation, we can eliminate the sliding phases $\phi_{x,y}$ by shifting the origin via continuous translation symmetry. In the basis $(|\bm{r},+\rangle,|\bm{r},-\rangle)$, the continuum Hamiltonian takes the following form, 
\begin{equation}
\begin{pmatrix}
h\left(-i\bm{\nabla}\right)&T(\bm{r})\\
T(\bm{r})&h\left(-i\bm{\nabla}\right)
\end{pmatrix}
\label{eq:continuum Ham}
\end{equation} 
where $h(\bm{p})$ and $T(\bm{r})$ is given by Eq.~\eqref{hp} and Eq.~\eqref{tr}. This continuum Hamiltonian will be the starting point for our following analysis.

{\bf Band structure:}
Eq.~\eqref{eq:continuum Ham} can be block-diagonalized using the eigenstates of the layer exchange operator $\sigma^x_{l,-l}$ \cite{tunablehubbardmodel}. In the $\sigma^x=\pm$ block sector, the reduced Hamiltonian is $H_{\sigma_x}\equiv h(-i\bm{\nabla})+\sigma_x T(\bm{r})$. Since all harmonics in the $T(\bm{r})$ expansion are of the form $e^{i2\sin(\theta/2)\bm{r}\cdot(\bm{Y}+\bm{G})\times\hat{z}}$, $T(\bm{r}+\bm{a}_{1m})=-T(\bm{r}),T(\bm{r}+\bm{a}_{2m})=-T(\bm{r})$ generally holds true for keeping arbitary orders of harmonic terms. Consequently, real-space wavefunctions of opposite $\sigma_x$ sectors are related by translations $\bm{a}_{1m}$ and $\bm{a}_{2m}$, allowing us to focus on the $\sigma_x=+$ sector. 

\begin{figure}[t]
\includegraphics[width=1.0\linewidth]{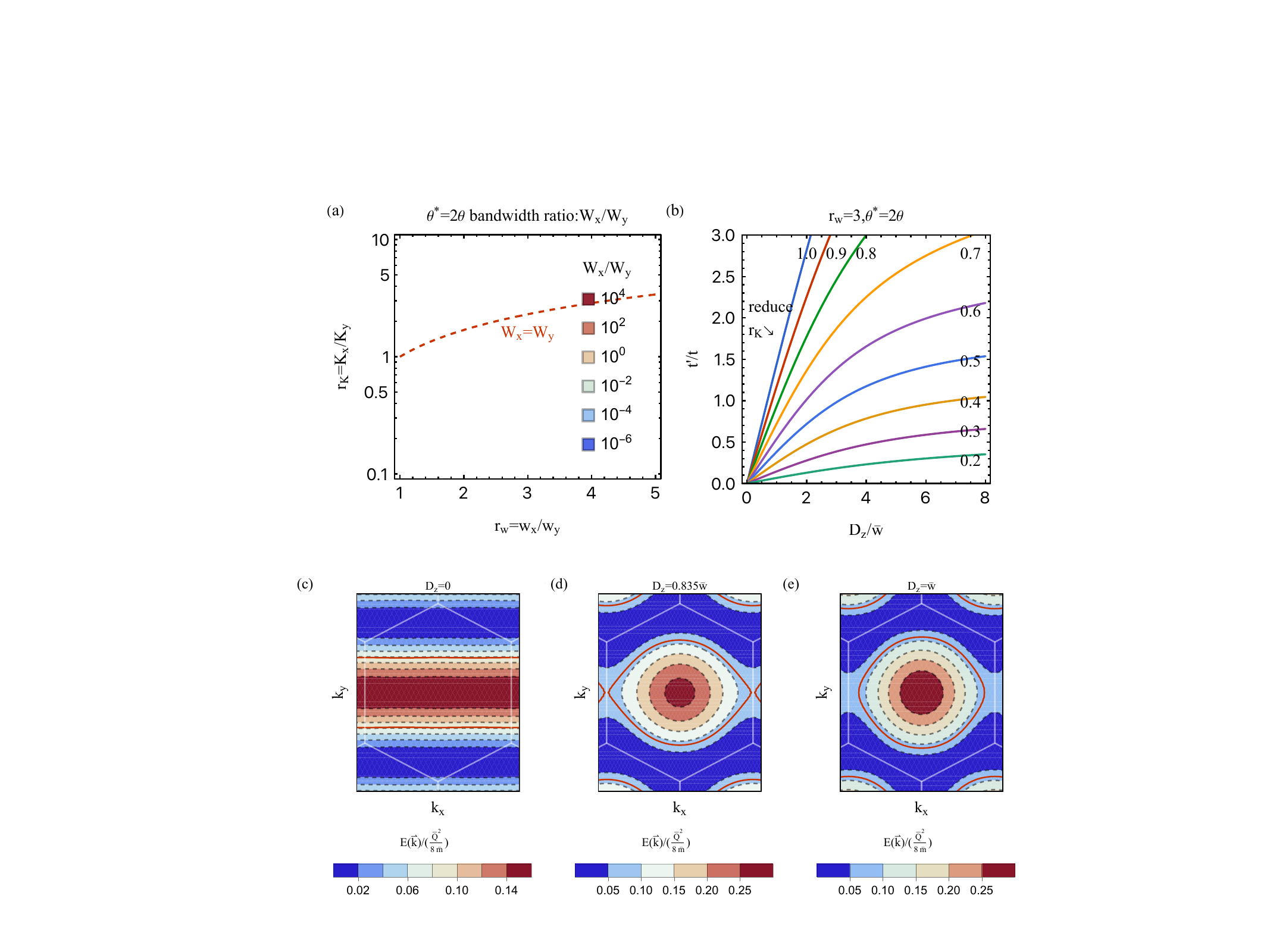}\caption{\label{fig:tuning}(a) In the absence of a displacement field and within the small twist angle limit, $r_w$ and $r_K$ determine the hopping direction of the anisotropic chains. (b) The displacement field $D_z$ allows tuning of the nearest neighbor interchain hopping $t'$. When $r_K$ is within a specific range, $t'$ can exceed the intrachain hopping $t$. (c) Quasi-1D Fermi surface at $D_z=0$. (d) At some particular $D_z$, the Fermi surface at half-filling features a Van Hove singularity. (e) Nearly isotropic Fermi surface at $D_z=\bar{w}$. In (c-e), the red lines depict the Fermi surfaces at half-filling. We measure $D_z$ in units of the interlayer tunneling energy scale $\bar{w}=\sqrt{w_xw_y}$.}
\end{figure}

In the two-harmonics approximation, $H_+$ can be simplified by decomposing into $H_x(x,-i\partial_x)+H_y(y,-i\partial_y)$ \cite{supp}. For small twist angles below a characteristic angle $\theta^*$, the top moiré band is isolated and the bandwidths $W_{x}$ of $H_{x}$ and $W_{y}$ of $H_{y}$ can differ significantly. We use the ratio $r_W=W_x/W_y$ to quantify band anisotropy. As shown in Fig.~\ref{fig:tuning}(a), the band anisotropy $r_W$ is completely determined by two dimensionless ratios at fixed twist angle $\theta$, $r_w=w_x/w_y$ (see $T(\bm{r})$ in Eq.~\eqref{tr}), and $r_K=(Q_x^2m_y)/(Q_y^2m_x)$, where $r_W$ is highly sensitive to $r_K$ and moderately sensitive to $r_w$. The highly anisotropic regime can be understood as parallel weakly coupled $x$- or $y$-directed chains. Importantly, as seen above, the $\sigma_x$ symmetry prohibits couplings between chains separated by $\bm{a}_{1m}$ or $\bm{a}_{2m}$. 

A practical way to break this $\sigma_x$ symmetry and induce neighboring interchain hoppings \cite{tunablehubbardmodel} is by adding the displacement field $(D_z/2)\sigma_z$. 
Here we compare the band structure changes under $D_z$ using representative parameters $(r_K=0.7,r_w=3,\theta=0.5\theta^*)$ in Fig.~\ref{fig:tuning} \cite{supp}. 
At $D_z=0$ (Fig.~\ref{fig:tuning}(c)), the dispersion is flat along $x$ direction and dispersive along $y$ direction, 
indicating quasi-1D $y$-directed chains. With $D_z$ increased, the non-dispersive nature along $x$ direction is lost. We can tune $D_z$ to make the half-filling Fermi surface pass through a single van Hove singularity at $\bm{Q}_x$ (Fig.~\ref{fig:tuning}(b)). For larger $D_z$, the Fermi surface can be nearly isotropic and enters 2D regime (Fig.~\ref{fig:tuning}(e)).

Across the $D_z$ range studied, the band structure can be well fitted using two dominant hoppings (see Fig.~\ref{fig:moire}(c)), $t$ (intrachain, along $\pm(\bm{a}_{2m}-\bm{a}_{1m})$) and $t'$ (interchain, along $\pm\bm{a}_{1m}$ and $\pm\bm{a}_{2m}$). This yields a effective tight-binding dispersion, 
\begin{equation}
    \epsilon(\bm{k})=2t\cos(\bm{k}\cdot(\bm{a}_{1m}-\bm{a}_{2m}))+2t'(\cos(\bm{k}\cdot\bm{a}_{1m})+\cos(\bm{k}\cdot\bm{a}_{2m}))
\end{equation}

We then investigate the tunability of $t'/t$ via $D_z$ for various $r_K$ values, as shown in Fig.~\ref{fig:tuning}(b). For materials with $0.5<r_K<1$, a moderate displacement field enables us to reach the $t'=t$ (and even $t'>t$) regime. These materials successfully realize tunable single-band triangular model in moiré superlattice systems, providing a promising platform for simulating one-dimensional quantum chains, frustrated anisotropic triangular lattices, and even square lattice (at $t'\gg t$ limit) on a single experimental system. 
For materials with $r_K>1$, the intrachain next-nearest-neighbor hopping $t''$ can appear and become much larger than $t$ for large $r_K$. The system is made up of $x-$directed chains. Applying the displacement field can also generate interchain hoppings \cite{supp}.

\begin{figure}
\includegraphics[width=1.0\linewidth]{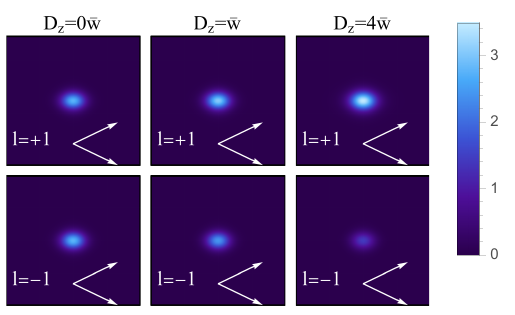}
\caption{\label{fig:wannierfunction}Real space structure of Wannier function at site $\bm{R}=0$. For illustrative purpose, the two primitive lattice vectors are translated by $\bm{a}_{1m}-\bm{a}_{2m}$. }
\end{figure}

{\bf Wannier function and interaction:}
To gain more insights into the nature of the twisted diamond bilayers from real space perspective, we construct localized Wannier functions \cite{supp} for the top moiré valence band using the same set of representative parameters $(r_K=0.7,r_w=3,\theta=0.5\theta^*)$ for various displacement fields $D_z$. 
Fig.~\ref{fig:wannierfunction} shows the layer-resolved charge density $\rho_l(\bm{r})=|W_{\bm{R}}(\bm{r},l)|^2$ of these Wannier functions $W_{\bm{R}}$.
At $D_z=0$, the charge is evenly distributed on each layer, whereas a nonzero $D_z$ induces a charge imbalance between two layers.

Having constructed the Wannier functions, the Coulomb interaction can be expanded onto this basis.  The leading term is the onsite Hubbard interaction, making the Hubbard model a natural framework for describing the correlation effects,

\begin{equation}
\label{hubbard_valence_band}
H=\sum_{i,j}t_{i,j}c_{i,s}^\dagger c_{j,s}+U\sum_{i}n_{i,\uparrow}n_{i,\downarrow}
\end{equation}
where $c_{i,s}$ ($c_{i,s}^\dagger$) annihilates (creates) a particle with spin $s$ at site $\bm{R}_i$.  $t_{i,j}$ equals $t$ for intrachain nearest-neighbor hopping and $t'$ for interchain nearest-neighbor hopping. $U$ is the strength of onsite Hubbard repulsion. Therefore, the twisted diamond bilayers with band extrema on $Y$ valley provides the first realization of displacement-field tuned anisotropy in a single-band triangular Hubbard model within moiré systems.

Moreover, it is worth emphasizing that, though $D_z\sim\bar{w}$ can lead to relatively large interchain hopping $t'$ (see Fig.~\ref{fig:tuning}(b)), the Wannier functions at $D_z=\bar{w}$ do not qualitatively change compared with the case of $D_z=0$. This implies that tuning the interchain hopping $t'$ via displacement field only has minimal impact on other quantities, such as intrachain hopping $t$ \cite{supp} and the Hubbard interaction $U$.

\begin{figure}[t]
\includegraphics[width=1.0\linewidth]{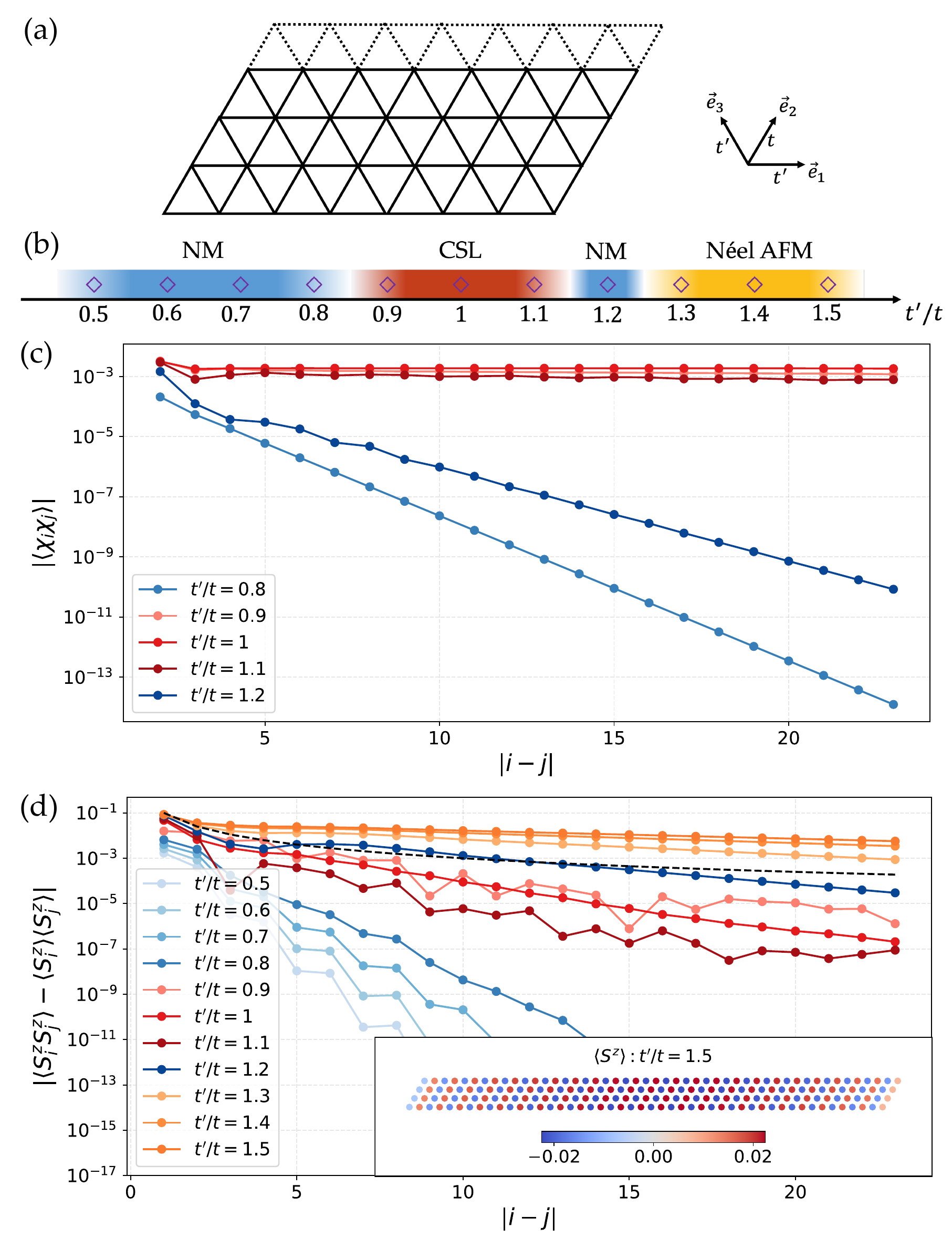}
\caption{(a) YC4 cylinder with periodic boundary along $\vec{e}_2$. Hopping anisotropy is introduced by setting $t$ along $\vec{e}_2$ and $t^\prime$ along  $\vec{e}_{1/3}$. (b) Ground state phase diagram of the anisotropic triangular lattice Hubbard model at half filling, with fixed interaction strength $U/t=10$. (c) Chiral correlation functions in the weakly anisotropic regime. (d) Spin correlation functions for various $t'/t$. The black dashed line represents a reference power law decay of $r^{-2}$. (Inset: the $\langle S^z\rangle$ distribution at $t^\prime/t=1.5$.)}\label{dmrg}
\end{figure}

{\bf DMRG results:}
After constructing the anisotropic triangular lattice Hubbard model, we employ DMRG to perform a preliminary investigation of the many-body physics at half filling, focusing on its evolution with hopping anisotropy $t'/t$ tuned by the displacement field. As illustrated in Fig.~\ref{dmrg}(a), we adopt the YC cylinder geometry with circumference $L_y=4$ and length up to $L_x=48$, and consider a symmetric implementation of hopping anisotropy (with $t$ along the $\vec{e}_2$ direction and $t^\prime$ along the symmetry equivalent $\vec{e}_{1}/\vec{e}_3$ directions). Our DMRG simulations explicitly preserve $U(1)_\text{charge} \times U(1)_\text{spin}$ symmetry, with bond dimensions up to $D=10000$.
The interaction strength is fixed at $U/t=10$ throughout this work, motivated by the discovery of CSL phase between a metal and a $120^\circ$ AFM phase in the isotropic limit~\cite{chiraltriangular,PhysRevB.110.L041113,PhysRevB.106.094420}. By tuning $t^\prime/t$ within $0.5 \leq t^\prime/t \leq 1.5$, we focus on the robustness of this CSL phase and the (magnetic) nature of other competing phases.

The ground state phase diagram as a function of $t^\prime/t$ for $U/t=10$ is shown in Fig.~\ref{dmrg}(b). We identify four phases within the range $0.5\leq t^\prime/t\leq 1.5$: the CSL phase at isotropic limit remains robust over $0.9\lesssim t^\prime/t\lesssim 1.1$, indicating its stability against moderate hopping anisotropy; two NM phases appear at larger hopping anisotropy $t^\prime/t\leq 0.8$ and $t^\prime/t\sim 1.2$; and a Néel AFM phase emerges for $t^\prime/t \gtrsim 1.3$, which can be smoothly connected to the square lattice limit $t^\prime/t\rightarrow\infty$.

To gain deeper insights into the nature of these phases, we first consider the chiral order characterized by the scalar chirality $\chi_i$, defined by:
\begin{equation}
    \chi_i = \vec{S}_i \cdot \left(\vec{S}_{i+\vec{e}_1} \times \vec{S}_{i+\vec{e}_2}\right)
\end{equation}
where $(i, i+\vec{e}_1, i+\vec{e}_2)$ denoting the three sites on the same upper triangle in counterclockwise order, and $\vec{S}_i$ is the spin operator at site $i$. The CSL phase is expected to have spontaneous scalar chirality $\langle\chi_i\rangle\neq 0$, indicating the breaking of time-reversal symmetry. However, due to the use of real number arithmetic in our DMRG simulations, the scalar chirality is constrained to vanish. Consequently, we evaluate the chiral correlation functions $\langle \chi_i \chi_j \rangle$ instead, as shown in Fig.~\ref{dmrg}(c). The presence of long-range chiral correlations within $0.9\lesssim t^\prime/t\lesssim 1.1$ provides strong evidence of spontaneous chiral order. Outside this regime, the chiral correlations exhibit fast exponential decay, indicating the absence of chiral order.

We then examine the behavior of spin correlation functions $\langle S_i^z S_j^z\rangle -\langle S_i^z \rangle\langle S_j^z\rangle$ to characterize the magnetic nature of the ground states. Due to the Mermin-Wagner theorem, true long-range magnetic order is forbidden in our quasi-1D DMRG simulations. The strongest possible spin correlations appear in a power law decaying form $r^{-\alpha}$, where $\alpha < 2$ implies a divergent zero-temperature spin susceptibility. The spin correlation functions (together with a reference line of $r^{-2}$ power law decay) are presented in Fig.~\ref{dmrg}(d). It is clear from the figure that the spin correlations exhibit fast exponential decay for $t^\prime/t \lesssim 1.2$, indicating the NM nature of this regime. Together with the chiral correlation results in Fig.~\ref{dmrg}(c), this suggests that the NM region is separated into two distinct (non-chiral) NM phases by an intervening CSL phase within $0.9\lesssim t^\prime/t\lesssim 1.1$. For $t^\prime/t\gtrsim 1.3$, spin correlations follow a power law decay much slower than $r^{-2}$, providing strong evidence for the development of magnetic ordering. The inset of Fig.~\ref{dmrg}(d) indicates that this magnetic ordered phase has clear Néel AFM pattern, with AFM correlations primarily along the strong $t^\prime$ bonds ($\vec{e}_{1/3}$ directions).

Our DMRG results suggest that tuning the hopping anisotropy $t^\prime/t$ in the triangular lattice Hubbard model can lead to rich physics at half filling. We expect that even more interesting physics, such as topological superconductivity~\cite{PhysRevLett.130.136003,PhysRevX.12.031009,PhysRevLett.125.157002,dopetj,PhysRevB.100.060506,dopehubbard,PhysRevLett.130.126001}, may arise upon doping away from half filling. We hope that our preliminary findings will stimulate further theoretical and experimental studies on the twisted diamond homobilayers with $Y$ valley.

{\bf Summary and Discussions:}
We study the continuum model of twisted Bravais diamond lattice homobilayers with band extrema at $Y$ valley. 
In the absence of a displacement field, the low-energy band structure is composed of decoupled chains. Introducing a displacement field generates interchain hopping, resulting in a single-band anisotropic triangular lattice Hubbard model with continuous tunability of the hopping anisotropy, offering a promising platform to study the anisotropy effects. We explore the interacting phase diagram by DMRG and find a chiral spin liquid phase, non-magnetic phases, and a Néel antiferromagnetic phase.

We conclude with several remarks. 
First, potential material candidates can be found using Ref. \cite{twistdatabase}. Using their notation, our setup corresponds to a rectangular lattice and layer groups that utilize the rectangular lattice as the conventional unit cell, rather than the primitive unit cell. We have identified $\text{Cu}_2\text{SO}_4$, $\text{Ag}_2\text{SeO}_4$, $\text{Au}_2\text{SO}_4$, $\text{Ag}_2\text{SO}_4$, among others, as suitable candidates. Second, we expect that including higher harmonics in the $T(\bm{r})$ expansion and the intralayer moiré potential $U(\bm{r})$ will not qualitatively change the top band orbitals or the displacement field's tunability of interchain hopping. The displacement field tunability relys on the layer exchange symmetry, which generally holds for arbitrary order harmonics. And within the small twist angle limit, the moiré potential $U(\bm{r})+\sigma_xT(\bm{r})$ in the $\sigma_x$ sector exhibits a deep profile, resulting in the confinement of orbitals around potential extrema. 
Third, the experimental signatures of QSLs, especially within moiré systems, are deserving of future investigative efforts \cite{PhysRevB.110.045116,PhysRevB.109.035127,PhysRevB.106.115108,PhysRevB.107.195102,PhysRevB.99.104425,PhysRevResearch.6.013043,Banerjee:2023aa}. And finally, if the displacement field can tune QSL and superconductors, nonuniform gating can create junctions between these phases \cite{PhysRevLett.131.146601,PhysRevB.105.L081405,PhysRevB.109.L041103,Anderson:2023aa}. These interfaces can host signatures of fractionalized electrons \cite{PhysRevB.35.8865,RevModPhys.78.17,PhysRevB.64.214511,Barkeshli_2014,Law:2017aa}.

{\bf Acknowledgement:} This work is supported in part by NSFC under Grant Nos. 12347107 and 12334003 (WS, CT, and HY), by MOSTC under Grant No. 2021YFA1400100 (HY), and by the New Cornerstone Science Foundation through the Xplorer Prize (HY).

\bibliography{main.bib}

\begin{thebibliography}{90}%
\makeatletter
\providecommand \@ifxundefined [1]{%
 \@ifx{#1\undefined}
}%
\providecommand \@ifnum [1]{%
 \ifnum #1\expandafter \@firstoftwo
 \else \expandafter \@secondoftwo
 \fi
}%
\providecommand \@ifx [1]{%
 \ifx #1\expandafter \@firstoftwo
 \else \expandafter \@secondoftwo
 \fi
}%
\providecommand \natexlab [1]{#1}%
\providecommand \enquote  [1]{``#1''}%
\providecommand \bibnamefont  [1]{#1}%
\providecommand \bibfnamefont [1]{#1}%
\providecommand \citenamefont [1]{#1}%
\providecommand \href@noop [0]{\@secondoftwo}%
\providecommand \href [0]{\begingroup \@sanitize@url \@href}%
\providecommand \@href[1]{\@@startlink{#1}\@@href}%
\providecommand \@@href[1]{\endgroup#1\@@endlink}%
\providecommand \@sanitize@url [0]{\catcode `\\12\catcode `\$12\catcode `\&12\catcode `\#12\catcode `\^12\catcode `\_12\catcode `\%12\relax}%
\providecommand \@@startlink[1]{}%
\providecommand \@@endlink[0]{}%
\providecommand \url  [0]{\begingroup\@sanitize@url \@url }%
\providecommand \@url [1]{\endgroup\@href {#1}{\urlprefix }}%
\providecommand \urlprefix  [0]{URL }%
\providecommand \Eprint [0]{\href }%
\providecommand \doibase [0]{https://doi.org/}%
\providecommand \selectlanguage [0]{\@gobble}%
\providecommand \bibinfo  [0]{\@secondoftwo}%
\providecommand \bibfield  [0]{\@secondoftwo}%
\providecommand \translation [1]{[#1]}%
\providecommand \BibitemOpen [0]{}%
\providecommand \bibitemStop [0]{}%
\providecommand \bibitemNoStop [0]{.\EOS\space}%
\providecommand \EOS [0]{\spacefactor3000\relax}%
\providecommand \BibitemShut  [1]{\csname bibitem#1\endcsname}%
\let\auto@bib@innerbib\@empty
\bibitem [{\citenamefont {Nuckolls}\ and\ \citenamefont {Yazdani}(2024)}]{mircroscopicofmoire}%
  \BibitemOpen
  \bibfield  {author} {\bibinfo {author} {\bibfnamefont {K.~P.}\ \bibnamefont {Nuckolls}}\ and\ \bibinfo {author} {\bibfnamefont {A.}~\bibnamefont {Yazdani}},\ }\bibfield  {title} {\bibinfo {title} {A microscopic perspective on moir{\'e} materials},\ }\href {https://doi.org/10.1038/s41578-024-00682-1} {\bibfield  {journal} {\bibinfo  {journal} {Nature Reviews Materials}\ }\textbf {\bibinfo {volume} {9}},\ \bibinfo {pages} {460} (\bibinfo {year} {2024})}\BibitemShut {NoStop}%
\bibitem [{\citenamefont {Kennes}\ \emph {et~al.}(2021)\citenamefont {Kennes}, \citenamefont {Claassen}, \citenamefont {Xian}, \citenamefont {Georges}, \citenamefont {Millis}, \citenamefont {Hone}, \citenamefont {Dean}, \citenamefont {Basov}, \citenamefont {Pasupathy},\ and\ \citenamefont {Rubio}}]{Kennes:2021aa}%
  \BibitemOpen
  \bibfield  {author} {\bibinfo {author} {\bibfnamefont {D.~M.}\ \bibnamefont {Kennes}}, \bibinfo {author} {\bibfnamefont {M.}~\bibnamefont {Claassen}}, \bibinfo {author} {\bibfnamefont {L.}~\bibnamefont {Xian}}, \bibinfo {author} {\bibfnamefont {A.}~\bibnamefont {Georges}}, \bibinfo {author} {\bibfnamefont {A.~J.}\ \bibnamefont {Millis}}, \bibinfo {author} {\bibfnamefont {J.}~\bibnamefont {Hone}}, \bibinfo {author} {\bibfnamefont {C.~R.}\ \bibnamefont {Dean}}, \bibinfo {author} {\bibfnamefont {D.~N.}\ \bibnamefont {Basov}}, \bibinfo {author} {\bibfnamefont {A.~N.}\ \bibnamefont {Pasupathy}},\ and\ \bibinfo {author} {\bibfnamefont {A.}~\bibnamefont {Rubio}},\ }\bibfield  {title} {\bibinfo {title} {Moir{\'e}heterostructures as a condensed-matter quantum simulator},\ }\href {https://doi.org/10.1038/s41567-020-01154-3} {\bibfield  {journal} {\bibinfo  {journal} {Nature Physics}\ }\textbf {\bibinfo {volume} {17}},\ \bibinfo {pages} {155} (\bibinfo {year} {2021})}\BibitemShut {NoStop}%
\bibitem [{\citenamefont {Andrei}\ \emph {et~al.}(2021)\citenamefont {Andrei}, \citenamefont {Efetov}, \citenamefont {Jarillo-Herrero}, \citenamefont {MacDonald}, \citenamefont {Mak}, \citenamefont {Senthil}, \citenamefont {Tutuc}, \citenamefont {Yazdani},\ and\ \citenamefont {Young}}]{reviewofmoiregraphene}%
  \BibitemOpen
  \bibfield  {author} {\bibinfo {author} {\bibfnamefont {E.~Y.}\ \bibnamefont {Andrei}}, \bibinfo {author} {\bibfnamefont {D.~K.}\ \bibnamefont {Efetov}}, \bibinfo {author} {\bibfnamefont {P.}~\bibnamefont {Jarillo-Herrero}}, \bibinfo {author} {\bibfnamefont {A.~H.}\ \bibnamefont {MacDonald}}, \bibinfo {author} {\bibfnamefont {K.~F.}\ \bibnamefont {Mak}}, \bibinfo {author} {\bibfnamefont {T.}~\bibnamefont {Senthil}}, \bibinfo {author} {\bibfnamefont {E.}~\bibnamefont {Tutuc}}, \bibinfo {author} {\bibfnamefont {A.}~\bibnamefont {Yazdani}},\ and\ \bibinfo {author} {\bibfnamefont {A.~F.}\ \bibnamefont {Young}},\ }\bibfield  {title} {\bibinfo {title} {The marvels of moir{\'e} materials},\ }\href {https://doi.org/10.1038/s41578-021-00284-1} {\bibfield  {journal} {\bibinfo  {journal} {Nature Reviews Materials}\ }\textbf {\bibinfo {volume} {6}},\ \bibinfo {pages} {201} (\bibinfo {year} {2021})}\BibitemShut {NoStop}%
\bibitem [{\citenamefont {Mak}\ and\ \citenamefont {Shan}(2022)}]{reviewofmoireTMD}%
  \BibitemOpen
  \bibfield  {author} {\bibinfo {author} {\bibfnamefont {K.~F.}\ \bibnamefont {Mak}}\ and\ \bibinfo {author} {\bibfnamefont {J.}~\bibnamefont {Shan}},\ }\bibfield  {title} {\bibinfo {title} {Semiconductor moir{\'e} materials},\ }\href {https://doi.org/10.1038/s41565-022-01165-6} {\bibfield  {journal} {\bibinfo  {journal} {Nature Nanotechnology}\ }\textbf {\bibinfo {volume} {17}},\ \bibinfo {pages} {686} (\bibinfo {year} {2022})}\BibitemShut {NoStop}%
\bibitem [{\citenamefont {Castellanos-Gomez}\ \emph {et~al.}(2022)\citenamefont {Castellanos-Gomez}, \citenamefont {Duan}, \citenamefont {Fei}, \citenamefont {Gutierrez}, \citenamefont {Huang}, \citenamefont {Huang}, \citenamefont {Quereda}, \citenamefont {Qian}, \citenamefont {Sutter},\ and\ \citenamefont {Sutter}}]{Castellanos-Gomez:2022aa}%
  \BibitemOpen
  \bibfield  {author} {\bibinfo {author} {\bibfnamefont {A.}~\bibnamefont {Castellanos-Gomez}}, \bibinfo {author} {\bibfnamefont {X.}~\bibnamefont {Duan}}, \bibinfo {author} {\bibfnamefont {Z.}~\bibnamefont {Fei}}, \bibinfo {author} {\bibfnamefont {H.~R.}\ \bibnamefont {Gutierrez}}, \bibinfo {author} {\bibfnamefont {Y.}~\bibnamefont {Huang}}, \bibinfo {author} {\bibfnamefont {X.}~\bibnamefont {Huang}}, \bibinfo {author} {\bibfnamefont {J.}~\bibnamefont {Quereda}}, \bibinfo {author} {\bibfnamefont {Q.}~\bibnamefont {Qian}}, \bibinfo {author} {\bibfnamefont {E.}~\bibnamefont {Sutter}},\ and\ \bibinfo {author} {\bibfnamefont {P.}~\bibnamefont {Sutter}},\ }\bibfield  {title} {\bibinfo {title} {Van der waals heterostructures},\ }\href {https://doi.org/10.1038/s43586-022-00139-1} {\bibfield  {journal} {\bibinfo  {journal} {Nature Reviews Methods Primers}\ }\textbf {\bibinfo {volume} {2}},\ \bibinfo {pages} {58} (\bibinfo {year} {2022})}\BibitemShut {NoStop}%
\bibitem [{\citenamefont {Lopes~dos Santos}\ \emph {et~al.}(2007)\citenamefont {Lopes~dos Santos}, \citenamefont {Peres},\ and\ \citenamefont {Castro~Neto}}]{moirebandstructureprl}%
  \BibitemOpen
  \bibfield  {author} {\bibinfo {author} {\bibfnamefont {J.~M.~B.}\ \bibnamefont {Lopes~dos Santos}}, \bibinfo {author} {\bibfnamefont {N.~M.~R.}\ \bibnamefont {Peres}},\ and\ \bibinfo {author} {\bibfnamefont {A.~H.}\ \bibnamefont {Castro~Neto}},\ }\bibfield  {title} {\bibinfo {title} {Graphene bilayer with a twist: Electronic structure},\ }\href {https://doi.org/10.1103/PhysRevLett.99.256802} {\bibfield  {journal} {\bibinfo  {journal} {Phys. Rev. Lett.}\ }\textbf {\bibinfo {volume} {99}},\ \bibinfo {pages} {256802} (\bibinfo {year} {2007})}\BibitemShut {NoStop}%
\bibitem [{\citenamefont {Bistritzer}\ and\ \citenamefont {MacDonald}(2011)}]{macdonaldmodel}%
  \BibitemOpen
  \bibfield  {author} {\bibinfo {author} {\bibfnamefont {R.}~\bibnamefont {Bistritzer}}\ and\ \bibinfo {author} {\bibfnamefont {A.~H.}\ \bibnamefont {MacDonald}},\ }\bibfield  {title} {\bibinfo {title} {Moir{\'e} bands in twisted double-layer graphene},\ }\href {https://doi.org/10.1073/pnas.1108174108} {\bibfield  {journal} {\bibinfo  {journal} {Proceedings of the National Academy of Sciences}\ }\textbf {\bibinfo {volume} {108}},\ \bibinfo {pages} {12233} (\bibinfo {year} {2011})},\ \Eprint {https://arxiv.org/abs/https://www.pnas.org/doi/pdf/10.1073/pnas.1108174108} {https://www.pnas.org/doi/pdf/10.1073/pnas.1108174108} \BibitemShut {NoStop}%
\bibitem [{\citenamefont {Cao}\ \emph {et~al.}(2018{\natexlab{a}})\citenamefont {Cao}, \citenamefont {Fatemi}, \citenamefont {Demir}, \citenamefont {Fang}, \citenamefont {Tomarken}, \citenamefont {Luo}, \citenamefont {Sanchez-Yamagishi}, \citenamefont {Watanabe}, \citenamefont {Taniguchi}, \citenamefont {Kaxiras}, \citenamefont {Ashoori},\ and\ \citenamefont {Jarillo-Herrero}}]{TBGinsulator}%
  \BibitemOpen
  \bibfield  {author} {\bibinfo {author} {\bibfnamefont {Y.}~\bibnamefont {Cao}}, \bibinfo {author} {\bibfnamefont {V.}~\bibnamefont {Fatemi}}, \bibinfo {author} {\bibfnamefont {A.}~\bibnamefont {Demir}}, \bibinfo {author} {\bibfnamefont {S.}~\bibnamefont {Fang}}, \bibinfo {author} {\bibfnamefont {S.~L.}\ \bibnamefont {Tomarken}}, \bibinfo {author} {\bibfnamefont {J.~Y.}\ \bibnamefont {Luo}}, \bibinfo {author} {\bibfnamefont {J.~D.}\ \bibnamefont {Sanchez-Yamagishi}}, \bibinfo {author} {\bibfnamefont {K.}~\bibnamefont {Watanabe}}, \bibinfo {author} {\bibfnamefont {T.}~\bibnamefont {Taniguchi}}, \bibinfo {author} {\bibfnamefont {E.}~\bibnamefont {Kaxiras}}, \bibinfo {author} {\bibfnamefont {R.~C.}\ \bibnamefont {Ashoori}},\ and\ \bibinfo {author} {\bibfnamefont {P.}~\bibnamefont {Jarillo-Herrero}},\ }\bibfield  {title} {\bibinfo {title} {Correlated insulator behaviour at half-filling in magic-angle graphene superlattices},\ }\href {https://doi.org/10.1038/nature26154} {\bibfield  {journal} {\bibinfo  {journal}
  {Nature}\ }\textbf {\bibinfo {volume} {556}},\ \bibinfo {pages} {80} (\bibinfo {year} {2018}{\natexlab{a}})}\BibitemShut {NoStop}%
\bibitem [{\citenamefont {Cao}\ \emph {et~al.}(2018{\natexlab{b}})\citenamefont {Cao}, \citenamefont {Fatemi}, \citenamefont {Fang}, \citenamefont {Watanabe}, \citenamefont {Taniguchi}, \citenamefont {Kaxiras},\ and\ \citenamefont {Jarillo-Herrero}}]{TBGsc}%
  \BibitemOpen
  \bibfield  {author} {\bibinfo {author} {\bibfnamefont {Y.}~\bibnamefont {Cao}}, \bibinfo {author} {\bibfnamefont {V.}~\bibnamefont {Fatemi}}, \bibinfo {author} {\bibfnamefont {S.}~\bibnamefont {Fang}}, \bibinfo {author} {\bibfnamefont {K.}~\bibnamefont {Watanabe}}, \bibinfo {author} {\bibfnamefont {T.}~\bibnamefont {Taniguchi}}, \bibinfo {author} {\bibfnamefont {E.}~\bibnamefont {Kaxiras}},\ and\ \bibinfo {author} {\bibfnamefont {P.}~\bibnamefont {Jarillo-Herrero}},\ }\bibfield  {title} {\bibinfo {title} {Unconventional superconductivity in magic-angle graphene superlattices},\ }\href {https://doi.org/10.1038/nature26160} {\bibfield  {journal} {\bibinfo  {journal} {Nature}\ }\textbf {\bibinfo {volume} {556}},\ \bibinfo {pages} {43} (\bibinfo {year} {2018}{\natexlab{b}})}\BibitemShut {NoStop}%
\bibitem [{\citenamefont {Lu}\ \emph {et~al.}(2019)\citenamefont {Lu}, \citenamefont {Stepanov}, \citenamefont {Yang}, \citenamefont {Xie}, \citenamefont {Aamir}, \citenamefont {Das}, \citenamefont {Urgell}, \citenamefont {Watanabe}, \citenamefont {Taniguchi}, \citenamefont {Zhang}, \citenamefont {Bachtold}, \citenamefont {MacDonald},\ and\ \citenamefont {Efetov}}]{TBGorbitalmagnet}%
  \BibitemOpen
  \bibfield  {author} {\bibinfo {author} {\bibfnamefont {X.}~\bibnamefont {Lu}}, \bibinfo {author} {\bibfnamefont {P.}~\bibnamefont {Stepanov}}, \bibinfo {author} {\bibfnamefont {W.}~\bibnamefont {Yang}}, \bibinfo {author} {\bibfnamefont {M.}~\bibnamefont {Xie}}, \bibinfo {author} {\bibfnamefont {M.~A.}\ \bibnamefont {Aamir}}, \bibinfo {author} {\bibfnamefont {I.}~\bibnamefont {Das}}, \bibinfo {author} {\bibfnamefont {C.}~\bibnamefont {Urgell}}, \bibinfo {author} {\bibfnamefont {K.}~\bibnamefont {Watanabe}}, \bibinfo {author} {\bibfnamefont {T.}~\bibnamefont {Taniguchi}}, \bibinfo {author} {\bibfnamefont {G.}~\bibnamefont {Zhang}}, \bibinfo {author} {\bibfnamefont {A.}~\bibnamefont {Bachtold}}, \bibinfo {author} {\bibfnamefont {A.~H.}\ \bibnamefont {MacDonald}},\ and\ \bibinfo {author} {\bibfnamefont {D.~K.}\ \bibnamefont {Efetov}},\ }\bibfield  {title} {\bibinfo {title} {Superconductors, orbital magnets and correlated states in magic-angle bilayer graphene},\ }\href
  {https://doi.org/10.1038/s41586-019-1695-0} {\bibfield  {journal} {\bibinfo  {journal} {Nature}\ }\textbf {\bibinfo {volume} {574}},\ \bibinfo {pages} {653} (\bibinfo {year} {2019})}\BibitemShut {NoStop}%
\bibitem [{\citenamefont {Jiang}\ \emph {et~al.}(2019)\citenamefont {Jiang}, \citenamefont {Lai}, \citenamefont {Watanabe}, \citenamefont {Taniguchi}, \citenamefont {Haule}, \citenamefont {Mao},\ and\ \citenamefont {Andrei}}]{nematicTBG2}%
  \BibitemOpen
  \bibfield  {author} {\bibinfo {author} {\bibfnamefont {Y.}~\bibnamefont {Jiang}}, \bibinfo {author} {\bibfnamefont {X.}~\bibnamefont {Lai}}, \bibinfo {author} {\bibfnamefont {K.}~\bibnamefont {Watanabe}}, \bibinfo {author} {\bibfnamefont {T.}~\bibnamefont {Taniguchi}}, \bibinfo {author} {\bibfnamefont {K.}~\bibnamefont {Haule}}, \bibinfo {author} {\bibfnamefont {J.}~\bibnamefont {Mao}},\ and\ \bibinfo {author} {\bibfnamefont {E.~Y.}\ \bibnamefont {Andrei}},\ }\bibfield  {title} {\bibinfo {title} {Charge order and broken rotational symmetry in magic-angle twisted bilayer graphene},\ }\href {https://doi.org/10.1038/s41586-019-1460-4} {\bibfield  {journal} {\bibinfo  {journal} {Nature}\ }\textbf {\bibinfo {volume} {573}},\ \bibinfo {pages} {91} (\bibinfo {year} {2019})}\BibitemShut {NoStop}%
\bibitem [{\citenamefont {Kerelsky}\ \emph {et~al.}(2019)\citenamefont {Kerelsky}, \citenamefont {McGilly}, \citenamefont {Kennes}, \citenamefont {Xian}, \citenamefont {Yankowitz}, \citenamefont {Chen}, \citenamefont {Watanabe}, \citenamefont {Taniguchi}, \citenamefont {Hone}, \citenamefont {Dean}, \citenamefont {Rubio},\ and\ \citenamefont {Pasupathy}}]{nematicTBG1}%
  \BibitemOpen
  \bibfield  {author} {\bibinfo {author} {\bibfnamefont {A.}~\bibnamefont {Kerelsky}}, \bibinfo {author} {\bibfnamefont {L.~J.}\ \bibnamefont {McGilly}}, \bibinfo {author} {\bibfnamefont {D.~M.}\ \bibnamefont {Kennes}}, \bibinfo {author} {\bibfnamefont {L.}~\bibnamefont {Xian}}, \bibinfo {author} {\bibfnamefont {M.}~\bibnamefont {Yankowitz}}, \bibinfo {author} {\bibfnamefont {S.}~\bibnamefont {Chen}}, \bibinfo {author} {\bibfnamefont {K.}~\bibnamefont {Watanabe}}, \bibinfo {author} {\bibfnamefont {T.}~\bibnamefont {Taniguchi}}, \bibinfo {author} {\bibfnamefont {J.}~\bibnamefont {Hone}}, \bibinfo {author} {\bibfnamefont {C.}~\bibnamefont {Dean}}, \bibinfo {author} {\bibfnamefont {A.}~\bibnamefont {Rubio}},\ and\ \bibinfo {author} {\bibfnamefont {A.~N.}\ \bibnamefont {Pasupathy}},\ }\bibfield  {title} {\bibinfo {title} {Maximized electron interactions at the magic angle in twisted bilayer graphene},\ }\href {https://doi.org/10.1038/s41586-019-1431-9} {\bibfield  {journal} {\bibinfo  {journal} {Nature}\ }\textbf
  {\bibinfo {volume} {572}},\ \bibinfo {pages} {95} (\bibinfo {year} {2019})}\BibitemShut {NoStop}%
\bibitem [{\citenamefont {Cao}\ \emph {et~al.}(2021)\citenamefont {Cao}, \citenamefont {Rodan-Legrain}, \citenamefont {Park}, \citenamefont {Yuan}, \citenamefont {Watanabe}, \citenamefont {Taniguchi}, \citenamefont {Fernandes}, \citenamefont {Fu},\ and\ \citenamefont {Jarillo-Herrero}}]{nematicscTBG}%
  \BibitemOpen
  \bibfield  {author} {\bibinfo {author} {\bibfnamefont {Y.}~\bibnamefont {Cao}}, \bibinfo {author} {\bibfnamefont {D.}~\bibnamefont {Rodan-Legrain}}, \bibinfo {author} {\bibfnamefont {J.~M.}\ \bibnamefont {Park}}, \bibinfo {author} {\bibfnamefont {N.~F.~Q.}\ \bibnamefont {Yuan}}, \bibinfo {author} {\bibfnamefont {K.}~\bibnamefont {Watanabe}}, \bibinfo {author} {\bibfnamefont {T.}~\bibnamefont {Taniguchi}}, \bibinfo {author} {\bibfnamefont {R.~M.}\ \bibnamefont {Fernandes}}, \bibinfo {author} {\bibfnamefont {L.}~\bibnamefont {Fu}},\ and\ \bibinfo {author} {\bibfnamefont {P.}~\bibnamefont {Jarillo-Herrero}},\ }\bibfield  {title} {\bibinfo {title} {Nematicity and competing orders in superconducting magic-angle graphene},\ }\href {https://doi.org/10.1126/science.abc2836} {\bibfield  {journal} {\bibinfo  {journal} {Science}\ }\textbf {\bibinfo {volume} {372}},\ \bibinfo {pages} {264} (\bibinfo {year} {2021})}\BibitemShut {NoStop}%
\bibitem [{\citenamefont {Nuckolls}\ \emph {et~al.}(2020)\citenamefont {Nuckolls}, \citenamefont {Oh}, \citenamefont {Wong}, \citenamefont {Lian}, \citenamefont {Watanabe}, \citenamefont {Taniguchi}, \citenamefont {Bernevig},\ and\ \citenamefont {Yazdani}}]{cherninsulatorTBG}%
  \BibitemOpen
  \bibfield  {author} {\bibinfo {author} {\bibfnamefont {K.~P.}\ \bibnamefont {Nuckolls}}, \bibinfo {author} {\bibfnamefont {M.}~\bibnamefont {Oh}}, \bibinfo {author} {\bibfnamefont {D.}~\bibnamefont {Wong}}, \bibinfo {author} {\bibfnamefont {B.}~\bibnamefont {Lian}}, \bibinfo {author} {\bibfnamefont {K.}~\bibnamefont {Watanabe}}, \bibinfo {author} {\bibfnamefont {T.}~\bibnamefont {Taniguchi}}, \bibinfo {author} {\bibfnamefont {B.~A.}\ \bibnamefont {Bernevig}},\ and\ \bibinfo {author} {\bibfnamefont {A.}~\bibnamefont {Yazdani}},\ }\bibfield  {title} {\bibinfo {title} {Strongly correlated chern insulators in magic-angle twisted bilayer graphene},\ }\href {https://doi.org/10.1038/s41586-020-3028-8} {\bibfield  {journal} {\bibinfo  {journal} {Nature}\ }\textbf {\bibinfo {volume} {588}},\ \bibinfo {pages} {610} (\bibinfo {year} {2020})}\BibitemShut {NoStop}%
\bibitem [{\citenamefont {Xie}\ \emph {et~al.}(2021)\citenamefont {Xie}, \citenamefont {Pierce}, \citenamefont {Park}, \citenamefont {Parker}, \citenamefont {Khalaf}, \citenamefont {Ledwith}, \citenamefont {Cao}, \citenamefont {Lee}, \citenamefont {Chen}, \citenamefont {Forrester}, \citenamefont {Watanabe}, \citenamefont {Taniguchi}, \citenamefont {Vishwanath}, \citenamefont {Jarillo-Herrero},\ and\ \citenamefont {Yacoby}}]{Xie:2021aa}%
  \BibitemOpen
  \bibfield  {author} {\bibinfo {author} {\bibfnamefont {Y.}~\bibnamefont {Xie}}, \bibinfo {author} {\bibfnamefont {A.~T.}\ \bibnamefont {Pierce}}, \bibinfo {author} {\bibfnamefont {J.~M.}\ \bibnamefont {Park}}, \bibinfo {author} {\bibfnamefont {D.~E.}\ \bibnamefont {Parker}}, \bibinfo {author} {\bibfnamefont {E.}~\bibnamefont {Khalaf}}, \bibinfo {author} {\bibfnamefont {P.}~\bibnamefont {Ledwith}}, \bibinfo {author} {\bibfnamefont {Y.}~\bibnamefont {Cao}}, \bibinfo {author} {\bibfnamefont {S.~H.}\ \bibnamefont {Lee}}, \bibinfo {author} {\bibfnamefont {S.}~\bibnamefont {Chen}}, \bibinfo {author} {\bibfnamefont {P.~R.}\ \bibnamefont {Forrester}}, \bibinfo {author} {\bibfnamefont {K.}~\bibnamefont {Watanabe}}, \bibinfo {author} {\bibfnamefont {T.}~\bibnamefont {Taniguchi}}, \bibinfo {author} {\bibfnamefont {A.}~\bibnamefont {Vishwanath}}, \bibinfo {author} {\bibfnamefont {P.}~\bibnamefont {Jarillo-Herrero}},\ and\ \bibinfo {author} {\bibfnamefont {A.}~\bibnamefont {Yacoby}},\ }\bibfield  {title} {\bibinfo
  {title} {Fractional chern insulators in magic-angle twisted bilayer graphene},\ }\href {https://doi.org/10.1038/s41586-021-04002-3} {\bibfield  {journal} {\bibinfo  {journal} {Nature}\ }\textbf {\bibinfo {volume} {600}},\ \bibinfo {pages} {439} (\bibinfo {year} {2021})}\BibitemShut {NoStop}%
\bibitem [{\citenamefont {Wu}\ \emph {et~al.}(2018)\citenamefont {Wu}, \citenamefont {Lovorn}, \citenamefont {Tutuc},\ and\ \citenamefont {MacDonald}}]{TMDhubbard}%
  \BibitemOpen
  \bibfield  {author} {\bibinfo {author} {\bibfnamefont {F.}~\bibnamefont {Wu}}, \bibinfo {author} {\bibfnamefont {T.}~\bibnamefont {Lovorn}}, \bibinfo {author} {\bibfnamefont {E.}~\bibnamefont {Tutuc}},\ and\ \bibinfo {author} {\bibfnamefont {A.~H.}\ \bibnamefont {MacDonald}},\ }\bibfield  {title} {\bibinfo {title} {Hubbard model physics in transition metal dichalcogenide moir\'e bands},\ }\href {https://doi.org/10.1103/PhysRevLett.121.026402} {\bibfield  {journal} {\bibinfo  {journal} {Phys. Rev. Lett.}\ }\textbf {\bibinfo {volume} {121}},\ \bibinfo {pages} {026402} (\bibinfo {year} {2018})}\BibitemShut {NoStop}%
\bibitem [{\citenamefont {Wu}\ \emph {et~al.}(2019)\citenamefont {Wu}, \citenamefont {Lovorn}, \citenamefont {Tutuc}, \citenamefont {Martin},\ and\ \citenamefont {MacDonald}}]{TMDTI}%
  \BibitemOpen
  \bibfield  {author} {\bibinfo {author} {\bibfnamefont {F.}~\bibnamefont {Wu}}, \bibinfo {author} {\bibfnamefont {T.}~\bibnamefont {Lovorn}}, \bibinfo {author} {\bibfnamefont {E.}~\bibnamefont {Tutuc}}, \bibinfo {author} {\bibfnamefont {I.}~\bibnamefont {Martin}},\ and\ \bibinfo {author} {\bibfnamefont {A.~H.}\ \bibnamefont {MacDonald}},\ }\bibfield  {title} {\bibinfo {title} {Topological insulators in twisted transition metal dichalcogenide homobilayers},\ }\href {https://doi.org/10.1103/PhysRevLett.122.086402} {\bibfield  {journal} {\bibinfo  {journal} {Phys. Rev. Lett.}\ }\textbf {\bibinfo {volume} {122}},\ \bibinfo {pages} {086402} (\bibinfo {year} {2019})}\BibitemShut {NoStop}%
\bibitem [{\citenamefont {Tang}\ \emph {et~al.}(2020)\citenamefont {Tang}, \citenamefont {Li}, \citenamefont {Li}, \citenamefont {Xu}, \citenamefont {Liu}, \citenamefont {Barmak}, \citenamefont {Watanabe}, \citenamefont {Taniguchi}, \citenamefont {MacDonald}, \citenamefont {Shan},\ and\ \citenamefont {Mak}}]{hubbardTMDexperiemnt}%
  \BibitemOpen
  \bibfield  {author} {\bibinfo {author} {\bibfnamefont {Y.}~\bibnamefont {Tang}}, \bibinfo {author} {\bibfnamefont {L.}~\bibnamefont {Li}}, \bibinfo {author} {\bibfnamefont {T.}~\bibnamefont {Li}}, \bibinfo {author} {\bibfnamefont {Y.}~\bibnamefont {Xu}}, \bibinfo {author} {\bibfnamefont {S.}~\bibnamefont {Liu}}, \bibinfo {author} {\bibfnamefont {K.}~\bibnamefont {Barmak}}, \bibinfo {author} {\bibfnamefont {K.}~\bibnamefont {Watanabe}}, \bibinfo {author} {\bibfnamefont {T.}~\bibnamefont {Taniguchi}}, \bibinfo {author} {\bibfnamefont {A.~H.}\ \bibnamefont {MacDonald}}, \bibinfo {author} {\bibfnamefont {J.}~\bibnamefont {Shan}},\ and\ \bibinfo {author} {\bibfnamefont {K.~F.}\ \bibnamefont {Mak}},\ }\bibfield  {title} {\bibinfo {title} {Simulation of hubbard model physics in wse2/ws2 moir{\'e}superlattices},\ }\href {https://doi.org/10.1038/s41586-020-2085-3} {\bibfield  {journal} {\bibinfo  {journal} {Nature}\ }\textbf {\bibinfo {volume} {579}},\ \bibinfo {pages} {353} (\bibinfo {year} {2020})}\BibitemShut
  {NoStop}%
\bibitem [{\citenamefont {Wang}\ \emph {et~al.}(2020)\citenamefont {Wang}, \citenamefont {Shih}, \citenamefont {Ghiotto}, \citenamefont {Xian}, \citenamefont {Rhodes}, \citenamefont {Tan}, \citenamefont {Claassen}, \citenamefont {Kennes}, \citenamefont {Bai}, \citenamefont {Kim}, \citenamefont {Watanabe}, \citenamefont {Taniguchi}, \citenamefont {Zhu}, \citenamefont {Hone}, \citenamefont {Rubio}, \citenamefont {Pasupathy},\ and\ \citenamefont {Dean}}]{correlatedinsulatorTMD}%
  \BibitemOpen
  \bibfield  {author} {\bibinfo {author} {\bibfnamefont {L.}~\bibnamefont {Wang}}, \bibinfo {author} {\bibfnamefont {E.-M.}\ \bibnamefont {Shih}}, \bibinfo {author} {\bibfnamefont {A.}~\bibnamefont {Ghiotto}}, \bibinfo {author} {\bibfnamefont {L.}~\bibnamefont {Xian}}, \bibinfo {author} {\bibfnamefont {D.~A.}\ \bibnamefont {Rhodes}}, \bibinfo {author} {\bibfnamefont {C.}~\bibnamefont {Tan}}, \bibinfo {author} {\bibfnamefont {M.}~\bibnamefont {Claassen}}, \bibinfo {author} {\bibfnamefont {D.~M.}\ \bibnamefont {Kennes}}, \bibinfo {author} {\bibfnamefont {Y.}~\bibnamefont {Bai}}, \bibinfo {author} {\bibfnamefont {B.}~\bibnamefont {Kim}}, \bibinfo {author} {\bibfnamefont {K.}~\bibnamefont {Watanabe}}, \bibinfo {author} {\bibfnamefont {T.}~\bibnamefont {Taniguchi}}, \bibinfo {author} {\bibfnamefont {X.}~\bibnamefont {Zhu}}, \bibinfo {author} {\bibfnamefont {J.}~\bibnamefont {Hone}}, \bibinfo {author} {\bibfnamefont {A.}~\bibnamefont {Rubio}}, \bibinfo {author} {\bibfnamefont {A.~N.}\ \bibnamefont {Pasupathy}},\
  and\ \bibinfo {author} {\bibfnamefont {C.~R.}\ \bibnamefont {Dean}},\ }\bibfield  {title} {\bibinfo {title} {Correlated electronic phases in twisted bilayer transition metal dichalcogenides},\ }\href {https://doi.org/10.1038/s41563-020-0708-6} {\bibfield  {journal} {\bibinfo  {journal} {Nature Materials}\ }\textbf {\bibinfo {volume} {19}},\ \bibinfo {pages} {861} (\bibinfo {year} {2020})}\BibitemShut {NoStop}%
\bibitem [{\citenamefont {Li}\ \emph {et~al.}(2021{\natexlab{a}})\citenamefont {Li}, \citenamefont {Jiang}, \citenamefont {Li}, \citenamefont {Zhang}, \citenamefont {Kang}, \citenamefont {Zhu}, \citenamefont {Watanabe}, \citenamefont {Taniguchi}, \citenamefont {Chowdhury}, \citenamefont {Fu}, \citenamefont {Shan},\ and\ \citenamefont {Mak}}]{MottTMD}%
  \BibitemOpen
  \bibfield  {author} {\bibinfo {author} {\bibfnamefont {T.}~\bibnamefont {Li}}, \bibinfo {author} {\bibfnamefont {S.}~\bibnamefont {Jiang}}, \bibinfo {author} {\bibfnamefont {L.}~\bibnamefont {Li}}, \bibinfo {author} {\bibfnamefont {Y.}~\bibnamefont {Zhang}}, \bibinfo {author} {\bibfnamefont {K.}~\bibnamefont {Kang}}, \bibinfo {author} {\bibfnamefont {J.}~\bibnamefont {Zhu}}, \bibinfo {author} {\bibfnamefont {K.}~\bibnamefont {Watanabe}}, \bibinfo {author} {\bibfnamefont {T.}~\bibnamefont {Taniguchi}}, \bibinfo {author} {\bibfnamefont {D.}~\bibnamefont {Chowdhury}}, \bibinfo {author} {\bibfnamefont {L.}~\bibnamefont {Fu}}, \bibinfo {author} {\bibfnamefont {J.}~\bibnamefont {Shan}},\ and\ \bibinfo {author} {\bibfnamefont {K.~F.}\ \bibnamefont {Mak}},\ }\bibfield  {title} {\bibinfo {title} {Continuous mott transition in semiconductor moir{\'e}superlattices},\ }\href {https://doi.org/10.1038/s41586-021-03853-0} {\bibfield  {journal} {\bibinfo  {journal} {Nature}\ }\textbf {\bibinfo {volume} {597}},\ \bibinfo
  {pages} {350} (\bibinfo {year} {2021}{\natexlab{a}})}\BibitemShut {NoStop}%
\bibitem [{\citenamefont {Xu}\ \emph {et~al.}(2020)\citenamefont {Xu}, \citenamefont {Liu}, \citenamefont {Rhodes}, \citenamefont {Watanabe}, \citenamefont {Taniguchi}, \citenamefont {Hone}, \citenamefont {Elser}, \citenamefont {Mak},\ and\ \citenamefont {Shan}}]{WignercrystalTMD1}%
  \BibitemOpen
  \bibfield  {author} {\bibinfo {author} {\bibfnamefont {Y.}~\bibnamefont {Xu}}, \bibinfo {author} {\bibfnamefont {S.}~\bibnamefont {Liu}}, \bibinfo {author} {\bibfnamefont {D.~A.}\ \bibnamefont {Rhodes}}, \bibinfo {author} {\bibfnamefont {K.}~\bibnamefont {Watanabe}}, \bibinfo {author} {\bibfnamefont {T.}~\bibnamefont {Taniguchi}}, \bibinfo {author} {\bibfnamefont {J.}~\bibnamefont {Hone}}, \bibinfo {author} {\bibfnamefont {V.}~\bibnamefont {Elser}}, \bibinfo {author} {\bibfnamefont {K.~F.}\ \bibnamefont {Mak}},\ and\ \bibinfo {author} {\bibfnamefont {J.}~\bibnamefont {Shan}},\ }\bibfield  {title} {\bibinfo {title} {Correlated insulating states at fractional fillings of moir{\'e}superlattices},\ }\href {https://doi.org/10.1038/s41586-020-2868-6} {\bibfield  {journal} {\bibinfo  {journal} {Nature}\ }\textbf {\bibinfo {volume} {587}},\ \bibinfo {pages} {214} (\bibinfo {year} {2020})}\BibitemShut {NoStop}%
\bibitem [{\citenamefont {Xu}\ \emph {et~al.}(2023)\citenamefont {Xu}, \citenamefont {Sun}, \citenamefont {Jia}, \citenamefont {Liu}, \citenamefont {Xu}, \citenamefont {Li}, \citenamefont {Gu}, \citenamefont {Watanabe}, \citenamefont {Taniguchi}, \citenamefont {Tong}, \citenamefont {Jia}, \citenamefont {Shi}, \citenamefont {Jiang}, \citenamefont {Zhang}, \citenamefont {Liu},\ and\ \citenamefont {Li}}]{chinaFQHE}%
  \BibitemOpen
  \bibfield  {author} {\bibinfo {author} {\bibfnamefont {F.}~\bibnamefont {Xu}}, \bibinfo {author} {\bibfnamefont {Z.}~\bibnamefont {Sun}}, \bibinfo {author} {\bibfnamefont {T.}~\bibnamefont {Jia}}, \bibinfo {author} {\bibfnamefont {C.}~\bibnamefont {Liu}}, \bibinfo {author} {\bibfnamefont {C.}~\bibnamefont {Xu}}, \bibinfo {author} {\bibfnamefont {C.}~\bibnamefont {Li}}, \bibinfo {author} {\bibfnamefont {Y.}~\bibnamefont {Gu}}, \bibinfo {author} {\bibfnamefont {K.}~\bibnamefont {Watanabe}}, \bibinfo {author} {\bibfnamefont {T.}~\bibnamefont {Taniguchi}}, \bibinfo {author} {\bibfnamefont {B.}~\bibnamefont {Tong}}, \bibinfo {author} {\bibfnamefont {J.}~\bibnamefont {Jia}}, \bibinfo {author} {\bibfnamefont {Z.}~\bibnamefont {Shi}}, \bibinfo {author} {\bibfnamefont {S.}~\bibnamefont {Jiang}}, \bibinfo {author} {\bibfnamefont {Y.}~\bibnamefont {Zhang}}, \bibinfo {author} {\bibfnamefont {X.}~\bibnamefont {Liu}},\ and\ \bibinfo {author} {\bibfnamefont {T.}~\bibnamefont {Li}},\ }\bibfield  {title} {\bibinfo {title}
  {Observation of integer and fractional quantum anomalous hall effects in twisted bilayer ${\mathrm{mote}}_{2}$},\ }\href {https://doi.org/10.1103/PhysRevX.13.031037} {\bibfield  {journal} {\bibinfo  {journal} {Phys. Rev. X}\ }\textbf {\bibinfo {volume} {13}},\ \bibinfo {pages} {031037} (\bibinfo {year} {2023})}\BibitemShut {NoStop}%
\bibitem [{\citenamefont {Li}\ \emph {et~al.}(2021{\natexlab{b}})\citenamefont {Li}, \citenamefont {Jiang}, \citenamefont {Shen}, \citenamefont {Zhang}, \citenamefont {Li}, \citenamefont {Tao}, \citenamefont {Devakul}, \citenamefont {Watanabe}, \citenamefont {Taniguchi}, \citenamefont {Fu}, \citenamefont {Shan},\ and\ \citenamefont {Mak}}]{QAHETMD}%
  \BibitemOpen
  \bibfield  {author} {\bibinfo {author} {\bibfnamefont {T.}~\bibnamefont {Li}}, \bibinfo {author} {\bibfnamefont {S.}~\bibnamefont {Jiang}}, \bibinfo {author} {\bibfnamefont {B.}~\bibnamefont {Shen}}, \bibinfo {author} {\bibfnamefont {Y.}~\bibnamefont {Zhang}}, \bibinfo {author} {\bibfnamefont {L.}~\bibnamefont {Li}}, \bibinfo {author} {\bibfnamefont {Z.}~\bibnamefont {Tao}}, \bibinfo {author} {\bibfnamefont {T.}~\bibnamefont {Devakul}}, \bibinfo {author} {\bibfnamefont {K.}~\bibnamefont {Watanabe}}, \bibinfo {author} {\bibfnamefont {T.}~\bibnamefont {Taniguchi}}, \bibinfo {author} {\bibfnamefont {L.}~\bibnamefont {Fu}}, \bibinfo {author} {\bibfnamefont {J.}~\bibnamefont {Shan}},\ and\ \bibinfo {author} {\bibfnamefont {K.~F.}\ \bibnamefont {Mak}},\ }\bibfield  {title} {\bibinfo {title} {Quantum anomalous hall effect from intertwined moir{\'e}bands},\ }\href {https://doi.org/10.1038/s41586-021-04171-1} {\bibfield  {journal} {\bibinfo  {journal} {Nature}\ }\textbf {\bibinfo {volume} {600}},\ \bibinfo {pages}
  {641} (\bibinfo {year} {2021}{\natexlab{b}})}\BibitemShut {NoStop}%
\bibitem [{\citenamefont {Zeng}\ \emph {et~al.}(2023)\citenamefont {Zeng}, \citenamefont {Xia}, \citenamefont {Kang}, \citenamefont {Zhu}, \citenamefont {Kn{\"u}ppel}, \citenamefont {Vaswani}, \citenamefont {Watanabe}, \citenamefont {Taniguchi}, \citenamefont {Mak},\ and\ \citenamefont {Shan}}]{FQAHETMD3}%
  \BibitemOpen
  \bibfield  {author} {\bibinfo {author} {\bibfnamefont {Y.}~\bibnamefont {Zeng}}, \bibinfo {author} {\bibfnamefont {Z.}~\bibnamefont {Xia}}, \bibinfo {author} {\bibfnamefont {K.}~\bibnamefont {Kang}}, \bibinfo {author} {\bibfnamefont {J.}~\bibnamefont {Zhu}}, \bibinfo {author} {\bibfnamefont {P.}~\bibnamefont {Kn{\"u}ppel}}, \bibinfo {author} {\bibfnamefont {C.}~\bibnamefont {Vaswani}}, \bibinfo {author} {\bibfnamefont {K.}~\bibnamefont {Watanabe}}, \bibinfo {author} {\bibfnamefont {T.}~\bibnamefont {Taniguchi}}, \bibinfo {author} {\bibfnamefont {K.~F.}\ \bibnamefont {Mak}},\ and\ \bibinfo {author} {\bibfnamefont {J.}~\bibnamefont {Shan}},\ }\bibfield  {title} {\bibinfo {title} {Thermodynamic evidence of fractional chern insulator in moir{\'e}mote2},\ }\href {https://doi.org/10.1038/s41586-023-06452-3} {\bibfield  {journal} {\bibinfo  {journal} {Nature}\ }\textbf {\bibinfo {volume} {622}},\ \bibinfo {pages} {69} (\bibinfo {year} {2023})}\BibitemShut {NoStop}%
\bibitem [{\citenamefont {Cai}\ \emph {et~al.}(2023)\citenamefont {Cai}, \citenamefont {Anderson}, \citenamefont {Wang}, \citenamefont {Zhang}, \citenamefont {Liu}, \citenamefont {Holtzmann}, \citenamefont {Zhang}, \citenamefont {Fan}, \citenamefont {Taniguchi}, \citenamefont {Watanabe}, \citenamefont {Ran}, \citenamefont {Cao}, \citenamefont {Fu}, \citenamefont {Xiao}, \citenamefont {Yao},\ and\ \citenamefont {Xu}}]{FQAHETMD2}%
  \BibitemOpen
  \bibfield  {author} {\bibinfo {author} {\bibfnamefont {J.}~\bibnamefont {Cai}}, \bibinfo {author} {\bibfnamefont {E.}~\bibnamefont {Anderson}}, \bibinfo {author} {\bibfnamefont {C.}~\bibnamefont {Wang}}, \bibinfo {author} {\bibfnamefont {X.}~\bibnamefont {Zhang}}, \bibinfo {author} {\bibfnamefont {X.}~\bibnamefont {Liu}}, \bibinfo {author} {\bibfnamefont {W.}~\bibnamefont {Holtzmann}}, \bibinfo {author} {\bibfnamefont {Y.}~\bibnamefont {Zhang}}, \bibinfo {author} {\bibfnamefont {F.}~\bibnamefont {Fan}}, \bibinfo {author} {\bibfnamefont {T.}~\bibnamefont {Taniguchi}}, \bibinfo {author} {\bibfnamefont {K.}~\bibnamefont {Watanabe}}, \bibinfo {author} {\bibfnamefont {Y.}~\bibnamefont {Ran}}, \bibinfo {author} {\bibfnamefont {T.}~\bibnamefont {Cao}}, \bibinfo {author} {\bibfnamefont {L.}~\bibnamefont {Fu}}, \bibinfo {author} {\bibfnamefont {D.}~\bibnamefont {Xiao}}, \bibinfo {author} {\bibfnamefont {W.}~\bibnamefont {Yao}},\ and\ \bibinfo {author} {\bibfnamefont {X.}~\bibnamefont {Xu}},\ }\bibfield  {title}
  {\bibinfo {title} {Signatures of fractional quantum anomalous hall states in twisted mote2},\ }\href {https://doi.org/10.1038/s41586-023-06289-w} {\bibfield  {journal} {\bibinfo  {journal} {Nature}\ }\textbf {\bibinfo {volume} {622}},\ \bibinfo {pages} {63} (\bibinfo {year} {2023})}\BibitemShut {NoStop}%
\bibitem [{\citenamefont {Park}\ \emph {et~al.}(2023)\citenamefont {Park}, \citenamefont {Cai}, \citenamefont {Anderson}, \citenamefont {Zhang}, \citenamefont {Zhu}, \citenamefont {Liu}, \citenamefont {Wang}, \citenamefont {Holtzmann}, \citenamefont {Hu}, \citenamefont {Liu}, \citenamefont {Taniguchi}, \citenamefont {Watanabe}, \citenamefont {Chu}, \citenamefont {Cao}, \citenamefont {Fu}, \citenamefont {Yao}, \citenamefont {Chang}, \citenamefont {Cobden}, \citenamefont {Xiao},\ and\ \citenamefont {Xu}}]{FQAHETMD1}%
  \BibitemOpen
  \bibfield  {author} {\bibinfo {author} {\bibfnamefont {H.}~\bibnamefont {Park}}, \bibinfo {author} {\bibfnamefont {J.}~\bibnamefont {Cai}}, \bibinfo {author} {\bibfnamefont {E.}~\bibnamefont {Anderson}}, \bibinfo {author} {\bibfnamefont {Y.}~\bibnamefont {Zhang}}, \bibinfo {author} {\bibfnamefont {J.}~\bibnamefont {Zhu}}, \bibinfo {author} {\bibfnamefont {X.}~\bibnamefont {Liu}}, \bibinfo {author} {\bibfnamefont {C.}~\bibnamefont {Wang}}, \bibinfo {author} {\bibfnamefont {W.}~\bibnamefont {Holtzmann}}, \bibinfo {author} {\bibfnamefont {C.}~\bibnamefont {Hu}}, \bibinfo {author} {\bibfnamefont {Z.}~\bibnamefont {Liu}}, \bibinfo {author} {\bibfnamefont {T.}~\bibnamefont {Taniguchi}}, \bibinfo {author} {\bibfnamefont {K.}~\bibnamefont {Watanabe}}, \bibinfo {author} {\bibfnamefont {J.-H.}\ \bibnamefont {Chu}}, \bibinfo {author} {\bibfnamefont {T.}~\bibnamefont {Cao}}, \bibinfo {author} {\bibfnamefont {L.}~\bibnamefont {Fu}}, \bibinfo {author} {\bibfnamefont {W.}~\bibnamefont {Yao}}, \bibinfo {author}
  {\bibfnamefont {C.-Z.}\ \bibnamefont {Chang}}, \bibinfo {author} {\bibfnamefont {D.}~\bibnamefont {Cobden}}, \bibinfo {author} {\bibfnamefont {D.}~\bibnamefont {Xiao}},\ and\ \bibinfo {author} {\bibfnamefont {X.}~\bibnamefont {Xu}},\ }\bibfield  {title} {\bibinfo {title} {Observation of fractionally quantized anomalous hall effect},\ }\href {https://doi.org/10.1038/s41586-023-06536-0} {\bibfield  {journal} {\bibinfo  {journal} {Nature}\ }\textbf {\bibinfo {volume} {622}},\ \bibinfo {pages} {74} (\bibinfo {year} {2023})}\BibitemShut {NoStop}%
\bibitem [{\citenamefont {Zhao}\ \emph {et~al.}(2023)\citenamefont {Zhao}, \citenamefont {Shen}, \citenamefont {Tao}, \citenamefont {Han}, \citenamefont {Kang}, \citenamefont {Watanabe}, \citenamefont {Taniguchi}, \citenamefont {Mak},\ and\ \citenamefont {Shan}}]{heavyfermionTMD}%
  \BibitemOpen
  \bibfield  {author} {\bibinfo {author} {\bibfnamefont {W.}~\bibnamefont {Zhao}}, \bibinfo {author} {\bibfnamefont {B.}~\bibnamefont {Shen}}, \bibinfo {author} {\bibfnamefont {Z.}~\bibnamefont {Tao}}, \bibinfo {author} {\bibfnamefont {Z.}~\bibnamefont {Han}}, \bibinfo {author} {\bibfnamefont {K.}~\bibnamefont {Kang}}, \bibinfo {author} {\bibfnamefont {K.}~\bibnamefont {Watanabe}}, \bibinfo {author} {\bibfnamefont {T.}~\bibnamefont {Taniguchi}}, \bibinfo {author} {\bibfnamefont {K.~F.}\ \bibnamefont {Mak}},\ and\ \bibinfo {author} {\bibfnamefont {J.}~\bibnamefont {Shan}},\ }\bibfield  {title} {\bibinfo {title} {Gate-tunable heavy fermions in a moir{\'e}kondo lattice},\ }\href {https://doi.org/10.1038/s41586-023-05800-7} {\bibfield  {journal} {\bibinfo  {journal} {Nature}\ }\textbf {\bibinfo {volume} {616}},\ \bibinfo {pages} {61} (\bibinfo {year} {2023})}\BibitemShut {NoStop}%
\bibitem [{\citenamefont {Kang}\ \emph {et~al.}(2024)\citenamefont {Kang}, \citenamefont {Shen}, \citenamefont {Qiu}, \citenamefont {Zeng}, \citenamefont {Xia}, \citenamefont {Watanabe}, \citenamefont {Taniguchi}, \citenamefont {Shan},\ and\ \citenamefont {Mak}}]{FQSHETMD}%
  \BibitemOpen
  \bibfield  {author} {\bibinfo {author} {\bibfnamefont {K.}~\bibnamefont {Kang}}, \bibinfo {author} {\bibfnamefont {B.}~\bibnamefont {Shen}}, \bibinfo {author} {\bibfnamefont {Y.}~\bibnamefont {Qiu}}, \bibinfo {author} {\bibfnamefont {Y.}~\bibnamefont {Zeng}}, \bibinfo {author} {\bibfnamefont {Z.}~\bibnamefont {Xia}}, \bibinfo {author} {\bibfnamefont {K.}~\bibnamefont {Watanabe}}, \bibinfo {author} {\bibfnamefont {T.}~\bibnamefont {Taniguchi}}, \bibinfo {author} {\bibfnamefont {J.}~\bibnamefont {Shan}},\ and\ \bibinfo {author} {\bibfnamefont {K.~F.}\ \bibnamefont {Mak}},\ }\bibfield  {title} {\bibinfo {title} {Evidence of the fractional quantum spin hall effect in moir{\'e}mote2},\ }\href {https://doi.org/10.1038/s41586-024-07214-5} {\bibfield  {journal} {\bibinfo  {journal} {Nature}\ }\textbf {\bibinfo {volume} {628}},\ \bibinfo {pages} {522} (\bibinfo {year} {2024})}\BibitemShut {NoStop}%
\bibitem [{\citenamefont {Guo}\ \emph {et~al.}(2025)\citenamefont {Guo}, \citenamefont {Pack}, \citenamefont {Swann}, \citenamefont {Holtzman}, \citenamefont {Cothrine}, \citenamefont {Watanabe}, \citenamefont {Taniguchi}, \citenamefont {Mandrus}, \citenamefont {Barmak}, \citenamefont {Hone}, \citenamefont {Millis}, \citenamefont {Pasupathy},\ and\ \citenamefont {Dean}}]{SCTMD2}%
  \BibitemOpen
  \bibfield  {author} {\bibinfo {author} {\bibfnamefont {Y.}~\bibnamefont {Guo}}, \bibinfo {author} {\bibfnamefont {J.}~\bibnamefont {Pack}}, \bibinfo {author} {\bibfnamefont {J.}~\bibnamefont {Swann}}, \bibinfo {author} {\bibfnamefont {L.}~\bibnamefont {Holtzman}}, \bibinfo {author} {\bibfnamefont {M.}~\bibnamefont {Cothrine}}, \bibinfo {author} {\bibfnamefont {K.}~\bibnamefont {Watanabe}}, \bibinfo {author} {\bibfnamefont {T.}~\bibnamefont {Taniguchi}}, \bibinfo {author} {\bibfnamefont {D.~G.}\ \bibnamefont {Mandrus}}, \bibinfo {author} {\bibfnamefont {K.}~\bibnamefont {Barmak}}, \bibinfo {author} {\bibfnamefont {J.}~\bibnamefont {Hone}}, \bibinfo {author} {\bibfnamefont {A.~J.}\ \bibnamefont {Millis}}, \bibinfo {author} {\bibfnamefont {A.}~\bibnamefont {Pasupathy}},\ and\ \bibinfo {author} {\bibfnamefont {C.~R.}\ \bibnamefont {Dean}},\ }\bibfield  {title} {\bibinfo {title} {Superconductivity in 5.0$\,^{\circ}$twisted bilayer wse2},\ }\href {https://doi.org/10.1038/s41586-024-08381-1} {\bibfield  {journal}
  {\bibinfo  {journal} {Nature}\ }\textbf {\bibinfo {volume} {637}},\ \bibinfo {pages} {839} (\bibinfo {year} {2025})}\BibitemShut {NoStop}%
\bibitem [{\citenamefont {Xia}\ \emph {et~al.}(2025)\citenamefont {Xia}, \citenamefont {Han}, \citenamefont {Watanabe}, \citenamefont {Taniguchi}, \citenamefont {Shan},\ and\ \citenamefont {Mak}}]{SCTMD1}%
  \BibitemOpen
  \bibfield  {author} {\bibinfo {author} {\bibfnamefont {Y.}~\bibnamefont {Xia}}, \bibinfo {author} {\bibfnamefont {Z.}~\bibnamefont {Han}}, \bibinfo {author} {\bibfnamefont {K.}~\bibnamefont {Watanabe}}, \bibinfo {author} {\bibfnamefont {T.}~\bibnamefont {Taniguchi}}, \bibinfo {author} {\bibfnamefont {J.}~\bibnamefont {Shan}},\ and\ \bibinfo {author} {\bibfnamefont {K.~F.}\ \bibnamefont {Mak}},\ }\bibfield  {title} {\bibinfo {title} {Superconductivity in twisted bilayer wse2},\ }\href {https://doi.org/10.1038/s41586-024-08116-2} {\bibfield  {journal} {\bibinfo  {journal} {Nature}\ }\textbf {\bibinfo {volume} {637}},\ \bibinfo {pages} {833} (\bibinfo {year} {2025})}\BibitemShut {NoStop}%
\bibitem [{\citenamefont {Geiser}\ \emph {et~al.}(1991)\citenamefont {Geiser}, \citenamefont {Wang}, \citenamefont {Carlson}, \citenamefont {Williams}, \citenamefont {Charlier}, \citenamefont {Heindl}, \citenamefont {Yaconi}, \citenamefont {Love},\ and\ \citenamefont {Lathrop}}]{Geiser:1991aa}%
  \BibitemOpen
  \bibfield  {author} {\bibinfo {author} {\bibfnamefont {U.}~\bibnamefont {Geiser}}, \bibinfo {author} {\bibfnamefont {H.~H.}\ \bibnamefont {Wang}}, \bibinfo {author} {\bibfnamefont {K.~D.}\ \bibnamefont {Carlson}}, \bibinfo {author} {\bibfnamefont {J.~M.}\ \bibnamefont {Williams}}, \bibinfo {author} {\bibfnamefont {H.~A.~J.}\ \bibnamefont {Charlier}}, \bibinfo {author} {\bibfnamefont {J.~E.}\ \bibnamefont {Heindl}}, \bibinfo {author} {\bibfnamefont {G.~A.}\ \bibnamefont {Yaconi}}, \bibinfo {author} {\bibfnamefont {B.~J.}\ \bibnamefont {Love}},\ and\ \bibinfo {author} {\bibfnamefont {M.~W.}\ \bibnamefont {Lathrop}},\ }\bibfield  {title} {\bibinfo {title} {Superconductivity at 2.8 k and 1.5 kbar in .kappa.-(bedt-ttf)2cu2(cn)3: the first organic superconductor containing a polymeric copper cyanide anion},\ }\href {https://doi.org/10.1021/ic00012a005} {\bibfield  {journal} {\bibinfo  {journal} {Inorganic Chemistry}\ }\textbf {\bibinfo {volume} {30}},\ \bibinfo {pages} {2586} (\bibinfo {year} {1991})}\BibitemShut
  {NoStop}%
\bibitem [{\citenamefont {Shimizu}\ \emph {et~al.}(2003)\citenamefont {Shimizu}, \citenamefont {Miyagawa}, \citenamefont {Kanoda}, \citenamefont {Maesato},\ and\ \citenamefont {Saito}}]{organicmott}%
  \BibitemOpen
  \bibfield  {author} {\bibinfo {author} {\bibfnamefont {Y.}~\bibnamefont {Shimizu}}, \bibinfo {author} {\bibfnamefont {K.}~\bibnamefont {Miyagawa}}, \bibinfo {author} {\bibfnamefont {K.}~\bibnamefont {Kanoda}}, \bibinfo {author} {\bibfnamefont {M.}~\bibnamefont {Maesato}},\ and\ \bibinfo {author} {\bibfnamefont {G.}~\bibnamefont {Saito}},\ }\bibfield  {title} {\bibinfo {title} {Spin liquid state in an organic mott insulator with a triangular lattice},\ }\href {https://doi.org/10.1103/PhysRevLett.91.107001} {\bibfield  {journal} {\bibinfo  {journal} {Phys. Rev. Lett.}\ }\textbf {\bibinfo {volume} {91}},\ \bibinfo {pages} {107001} (\bibinfo {year} {2003})}\BibitemShut {NoStop}%
\bibitem [{\citenamefont {Itou}\ \emph {et~al.}(2008)\citenamefont {Itou}, \citenamefont {Oyamada}, \citenamefont {Maegawa}, \citenamefont {Tamura},\ and\ \citenamefont {Kato}}]{Etcompound}%
  \BibitemOpen
  \bibfield  {author} {\bibinfo {author} {\bibfnamefont {T.}~\bibnamefont {Itou}}, \bibinfo {author} {\bibfnamefont {A.}~\bibnamefont {Oyamada}}, \bibinfo {author} {\bibfnamefont {S.}~\bibnamefont {Maegawa}}, \bibinfo {author} {\bibfnamefont {M.}~\bibnamefont {Tamura}},\ and\ \bibinfo {author} {\bibfnamefont {R.}~\bibnamefont {Kato}},\ }\bibfield  {title} {\bibinfo {title} {Quantum spin liquid in the spin-$1∕2$ triangular antiferromagnet $\mathrm{Et}{\mathrm{me}}_{3}\mathrm{Sb}{[\mathrm{Pd}{(\text{dmit})}_{2}]}_{2}$},\ }\href {https://doi.org/10.1103/PhysRevB.77.104413} {\bibfield  {journal} {\bibinfo  {journal} {Phys. Rev. B}\ }\textbf {\bibinfo {volume} {77}},\ \bibinfo {pages} {104413} (\bibinfo {year} {2008})}\BibitemShut {NoStop}%
\bibitem [{\citenamefont {Itou}\ \emph {et~al.}(2009)\citenamefont {Itou}, \citenamefont {Oyamada}, \citenamefont {Maegawa}, \citenamefont {Tamura},\ and\ \citenamefont {Kato}}]{NMRET}%
  \BibitemOpen
  \bibfield  {author} {\bibinfo {author} {\bibfnamefont {T.}~\bibnamefont {Itou}}, \bibinfo {author} {\bibfnamefont {A.}~\bibnamefont {Oyamada}}, \bibinfo {author} {\bibfnamefont {S.}~\bibnamefont {Maegawa}}, \bibinfo {author} {\bibfnamefont {M.}~\bibnamefont {Tamura}},\ and\ \bibinfo {author} {\bibfnamefont {R.}~\bibnamefont {Kato}},\ }\bibfield  {title} {\bibinfo {title} {13c nmr study of the spin-liquid state in the triangular quantum antiferromagnet etme3sb{$[$}pd(dmit)2{$]$}2},\ }\href {https://doi.org/10.1088/1742-6596/145/1/012039} {\bibfield  {journal} {\bibinfo  {journal} {Journal of Physics: Conference Series}\ }\textbf {\bibinfo {volume} {145}},\ \bibinfo {pages} {012039} (\bibinfo {year} {2009})}\BibitemShut {NoStop}%
\bibitem [{\citenamefont {Kanoda}\ and\ \citenamefont {Kato}(2011)}]{annurevorganic}%
  \BibitemOpen
  \bibfield  {author} {\bibinfo {author} {\bibfnamefont {K.}~\bibnamefont {Kanoda}}\ and\ \bibinfo {author} {\bibfnamefont {R.}~\bibnamefont {Kato}},\ }\bibfield  {title} {\bibinfo {title} {Mott physics in organic conductors with triangular lattices},\ }\href {https://doi.org/https://doi.org/10.1146/annurev-conmatphys-062910-140521} {\bibfield  {journal} {\bibinfo  {journal} {Annual Review of Condensed Matter Physics}\ }\textbf {\bibinfo {volume} {2}},\ \bibinfo {pages} {167} (\bibinfo {year} {2011})}\BibitemShut {NoStop}%
\bibitem [{\citenamefont {Powell}\ and\ \citenamefont {McKenzie}(2011)}]{IOPrevieworganic}%
  \BibitemOpen
  \bibfield  {author} {\bibinfo {author} {\bibfnamefont {B.~J.}\ \bibnamefont {Powell}}\ and\ \bibinfo {author} {\bibfnamefont {R.~H.}\ \bibnamefont {McKenzie}},\ }\bibfield  {title} {\bibinfo {title} {Quantum frustration in organic mott insulators: from spin liquids to unconventional superconductors},\ }\href {https://doi.org/10.1088/0034-4885/74/5/056501} {\bibfield  {journal} {\bibinfo  {journal} {Reports on Progress in Physics}\ }\textbf {\bibinfo {volume} {74}},\ \bibinfo {pages} {056501} (\bibinfo {year} {2011})}\BibitemShut {NoStop}%
\bibitem [{\citenamefont {Yamashita}\ \emph {et~al.}(2008)\citenamefont {Yamashita}, \citenamefont {Nakazawa}, \citenamefont {Oguni}, \citenamefont {Oshima}, \citenamefont {Nojiri}, \citenamefont {Shimizu}, \citenamefont {Miyagawa},\ and\ \citenamefont {Kanoda}}]{Yamashita:2008aa}%
  \BibitemOpen
  \bibfield  {author} {\bibinfo {author} {\bibfnamefont {S.}~\bibnamefont {Yamashita}}, \bibinfo {author} {\bibfnamefont {Y.}~\bibnamefont {Nakazawa}}, \bibinfo {author} {\bibfnamefont {M.}~\bibnamefont {Oguni}}, \bibinfo {author} {\bibfnamefont {Y.}~\bibnamefont {Oshima}}, \bibinfo {author} {\bibfnamefont {H.}~\bibnamefont {Nojiri}}, \bibinfo {author} {\bibfnamefont {Y.}~\bibnamefont {Shimizu}}, \bibinfo {author} {\bibfnamefont {K.}~\bibnamefont {Miyagawa}},\ and\ \bibinfo {author} {\bibfnamefont {K.}~\bibnamefont {Kanoda}},\ }\bibfield  {title} {\bibinfo {title} {Thermodynamic properties of a spin-1/2 spin-liquid state in a κ-type organic salt},\ }\href {https://doi.org/10.1038/nphys942} {\bibfield  {journal} {\bibinfo  {journal} {Nature Physics}\ }\textbf {\bibinfo {volume} {4}},\ \bibinfo {pages} {459} (\bibinfo {year} {2008})}\BibitemShut {NoStop}%
\bibitem [{\citenamefont {Yamashita}\ \emph {et~al.}(2010)\citenamefont {Yamashita}, \citenamefont {Nakata}, \citenamefont {Senshu}, \citenamefont {Nagata}, \citenamefont {Yamamoto}, \citenamefont {Kato}, \citenamefont {Shibauchi},\ and\ \citenamefont {Matsuda}}]{Yamashita:2010aa}%
  \BibitemOpen
  \bibfield  {author} {\bibinfo {author} {\bibfnamefont {M.}~\bibnamefont {Yamashita}}, \bibinfo {author} {\bibfnamefont {N.}~\bibnamefont {Nakata}}, \bibinfo {author} {\bibfnamefont {Y.}~\bibnamefont {Senshu}}, \bibinfo {author} {\bibfnamefont {M.}~\bibnamefont {Nagata}}, \bibinfo {author} {\bibfnamefont {H.~M.}\ \bibnamefont {Yamamoto}}, \bibinfo {author} {\bibfnamefont {R.}~\bibnamefont {Kato}}, \bibinfo {author} {\bibfnamefont {T.}~\bibnamefont {Shibauchi}},\ and\ \bibinfo {author} {\bibfnamefont {Y.}~\bibnamefont {Matsuda}},\ }\bibfield  {title} {\bibinfo {title} {Highly mobile gapless excitations in a two-dimensional candidate quantum spin liquid},\ }\href {https://doi.org/10.1126/science.1188200} {\bibfield  {journal} {\bibinfo  {journal} {Science}\ }\textbf {\bibinfo {volume} {328}},\ \bibinfo {pages} {1246} (\bibinfo {year} {2010})}\BibitemShut {NoStop}%
\bibitem [{\citenamefont {Hassan}\ \emph {et~al.}(2018)\citenamefont {Hassan}, \citenamefont {Cunningham}, \citenamefont {Mourigal}, \citenamefont {Zhilyaeva}, \citenamefont {Torunova}, \citenamefont {Lyubovskaya}, \citenamefont {Schlueter},\ and\ \citenamefont {Drichko}}]{Hassan:2018aa}%
  \BibitemOpen
  \bibfield  {author} {\bibinfo {author} {\bibfnamefont {N.}~\bibnamefont {Hassan}}, \bibinfo {author} {\bibfnamefont {S.}~\bibnamefont {Cunningham}}, \bibinfo {author} {\bibfnamefont {M.}~\bibnamefont {Mourigal}}, \bibinfo {author} {\bibfnamefont {E.~I.}\ \bibnamefont {Zhilyaeva}}, \bibinfo {author} {\bibfnamefont {S.~A.}\ \bibnamefont {Torunova}}, \bibinfo {author} {\bibfnamefont {R.~N.}\ \bibnamefont {Lyubovskaya}}, \bibinfo {author} {\bibfnamefont {J.~A.}\ \bibnamefont {Schlueter}},\ and\ \bibinfo {author} {\bibfnamefont {N.}~\bibnamefont {Drichko}},\ }\bibfield  {title} {\bibinfo {title} {Evidence for a quantum dipole liquid state in an organic quasi--two-dimensional material},\ }\href {https://doi.org/10.1126/science.aan6286} {\bibfield  {journal} {\bibinfo  {journal} {Science}\ }\textbf {\bibinfo {volume} {360}},\ \bibinfo {pages} {1101} (\bibinfo {year} {2018})}\BibitemShut {NoStop}%
\bibitem [{\citenamefont {Bourgeois-Hope}\ \emph {et~al.}(2019)\citenamefont {Bourgeois-Hope}, \citenamefont {Lalibert\'e}, \citenamefont {Lefran\ifmmode~\mbox{\c{c}}\else \c{c}\fi{}ois}, \citenamefont {Grissonnanche}, \citenamefont {de~Cotret}, \citenamefont {Gordon}, \citenamefont {Kitou}, \citenamefont {Sawa}, \citenamefont {Cui}, \citenamefont {Kato}, \citenamefont {Taillefer},\ and\ \citenamefont {Doiron-Leyraud}}]{PhysRevX.9.041051}%
  \BibitemOpen
  \bibfield  {author} {\bibinfo {author} {\bibfnamefont {P.}~\bibnamefont {Bourgeois-Hope}}, \bibinfo {author} {\bibfnamefont {F.}~\bibnamefont {Lalibert\'e}}, \bibinfo {author} {\bibfnamefont {E.}~\bibnamefont {Lefran\ifmmode~\mbox{\c{c}}\else \c{c}\fi{}ois}}, \bibinfo {author} {\bibfnamefont {G.}~\bibnamefont {Grissonnanche}}, \bibinfo {author} {\bibfnamefont {S.~R.}\ \bibnamefont {de~Cotret}}, \bibinfo {author} {\bibfnamefont {R.}~\bibnamefont {Gordon}}, \bibinfo {author} {\bibfnamefont {S.}~\bibnamefont {Kitou}}, \bibinfo {author} {\bibfnamefont {H.}~\bibnamefont {Sawa}}, \bibinfo {author} {\bibfnamefont {H.}~\bibnamefont {Cui}}, \bibinfo {author} {\bibfnamefont {R.}~\bibnamefont {Kato}}, \bibinfo {author} {\bibfnamefont {L.}~\bibnamefont {Taillefer}},\ and\ \bibinfo {author} {\bibfnamefont {N.}~\bibnamefont {Doiron-Leyraud}},\ }\bibfield  {title} {\bibinfo {title} {Thermal conductivity of the quantum spin liquid candidate
  ${\mathrm{etme}}_{3}\mathrm{Sb}\mathbf{[}\mathrm{Pd}\mathbf{(}\mathrm{dmit}{\mathbf{)}}_{2}{\mathbf{]}}_{2}$: No evidence of mobile gapless excitations},\ }\href {https://doi.org/10.1103/PhysRevX.9.041051} {\bibfield  {journal} {\bibinfo  {journal} {Phys. Rev. X}\ }\textbf {\bibinfo {volume} {9}},\ \bibinfo {pages} {041051} (\bibinfo {year} {2019})}\BibitemShut {NoStop}%
\bibitem [{\citenamefont {Ni}\ \emph {et~al.}(2019)\citenamefont {Ni}, \citenamefont {Pan}, \citenamefont {Song}, \citenamefont {Huang}, \citenamefont {Zeng}, \citenamefont {Yu}, \citenamefont {Cheng}, \citenamefont {Wang}, \citenamefont {Dai}, \citenamefont {Kato},\ and\ \citenamefont {Li}}]{PhysRevLett.123.247204}%
  \BibitemOpen
  \bibfield  {author} {\bibinfo {author} {\bibfnamefont {J.~M.}\ \bibnamefont {Ni}}, \bibinfo {author} {\bibfnamefont {B.~L.}\ \bibnamefont {Pan}}, \bibinfo {author} {\bibfnamefont {B.~Q.}\ \bibnamefont {Song}}, \bibinfo {author} {\bibfnamefont {Y.~Y.}\ \bibnamefont {Huang}}, \bibinfo {author} {\bibfnamefont {J.~Y.}\ \bibnamefont {Zeng}}, \bibinfo {author} {\bibfnamefont {Y.~J.}\ \bibnamefont {Yu}}, \bibinfo {author} {\bibfnamefont {E.~J.}\ \bibnamefont {Cheng}}, \bibinfo {author} {\bibfnamefont {L.~S.}\ \bibnamefont {Wang}}, \bibinfo {author} {\bibfnamefont {D.~Z.}\ \bibnamefont {Dai}}, \bibinfo {author} {\bibfnamefont {R.}~\bibnamefont {Kato}},\ and\ \bibinfo {author} {\bibfnamefont {S.~Y.}\ \bibnamefont {Li}},\ }\bibfield  {title} {\bibinfo {title} {Absence of magnetic thermal conductivity in the quantum spin liquid candidate ${\mathrm{etme}}_{3}\mathrm{Sb}[\mathrm{Pd}(\text{dmit}{)}_{2}{]}_{2}$},\ }\href {https://doi.org/10.1103/PhysRevLett.123.247204} {\bibfield  {journal} {\bibinfo  {journal} {Phys.
  Rev. Lett.}\ }\textbf {\bibinfo {volume} {123}},\ \bibinfo {pages} {247204} (\bibinfo {year} {2019})}\BibitemShut {NoStop}%
\bibitem [{\citenamefont {Watanabe}\ \emph {et~al.}(2012)\citenamefont {Watanabe}, \citenamefont {Yamashita}, \citenamefont {Tonegawa}, \citenamefont {Oshima}, \citenamefont {Yamamoto}, \citenamefont {Kato}, \citenamefont {Sheikin}, \citenamefont {Behnia}, \citenamefont {Terashima}, \citenamefont {Uji}, \citenamefont {Shibauchi},\ and\ \citenamefont {Matsuda}}]{Watanabe:2012aa}%
  \BibitemOpen
  \bibfield  {author} {\bibinfo {author} {\bibfnamefont {D.}~\bibnamefont {Watanabe}}, \bibinfo {author} {\bibfnamefont {M.}~\bibnamefont {Yamashita}}, \bibinfo {author} {\bibfnamefont {S.}~\bibnamefont {Tonegawa}}, \bibinfo {author} {\bibfnamefont {Y.}~\bibnamefont {Oshima}}, \bibinfo {author} {\bibfnamefont {H.~M.}\ \bibnamefont {Yamamoto}}, \bibinfo {author} {\bibfnamefont {R.}~\bibnamefont {Kato}}, \bibinfo {author} {\bibfnamefont {I.}~\bibnamefont {Sheikin}}, \bibinfo {author} {\bibfnamefont {K.}~\bibnamefont {Behnia}}, \bibinfo {author} {\bibfnamefont {T.}~\bibnamefont {Terashima}}, \bibinfo {author} {\bibfnamefont {S.}~\bibnamefont {Uji}}, \bibinfo {author} {\bibfnamefont {T.}~\bibnamefont {Shibauchi}},\ and\ \bibinfo {author} {\bibfnamefont {Y.}~\bibnamefont {Matsuda}},\ }\bibfield  {title} {\bibinfo {title} {Novel pauli-paramagnetic quantum phase in a mott insulator},\ }\href {https://doi.org/10.1038/ncomms2082} {\bibfield  {journal} {\bibinfo  {journal} {Nature Communications}\ }\textbf {\bibinfo
  {volume} {3}},\ \bibinfo {pages} {1090} (\bibinfo {year} {2012})}\BibitemShut {NoStop}%
\bibitem [{\citenamefont {Szasz}\ \emph {et~al.}(2020)\citenamefont {Szasz}, \citenamefont {Motruk}, \citenamefont {Zaletel},\ and\ \citenamefont {Moore}}]{chiraltriangular}%
  \BibitemOpen
  \bibfield  {author} {\bibinfo {author} {\bibfnamefont {A.}~\bibnamefont {Szasz}}, \bibinfo {author} {\bibfnamefont {J.}~\bibnamefont {Motruk}}, \bibinfo {author} {\bibfnamefont {M.~P.}\ \bibnamefont {Zaletel}},\ and\ \bibinfo {author} {\bibfnamefont {J.~E.}\ \bibnamefont {Moore}},\ }\bibfield  {title} {\bibinfo {title} {Chiral spin liquid phase of the triangular lattice hubbard model: A density matrix renormalization group study},\ }\href {https://doi.org/10.1103/PhysRevX.10.021042} {\bibfield  {journal} {\bibinfo  {journal} {Phys. Rev. X}\ }\textbf {\bibinfo {volume} {10}},\ \bibinfo {pages} {021042} (\bibinfo {year} {2020})}\BibitemShut {NoStop}%
\bibitem [{\citenamefont {Zhu}(2024)}]{PhysRevB.110.L041113}%
  \BibitemOpen
  \bibfield  {author} {\bibinfo {author} {\bibfnamefont {Z.}~\bibnamefont {Zhu}},\ }\bibfield  {title} {\bibinfo {title} {Chiral spin liquid versus mott antiferromagnetism in the triangular-lattice hubbard model},\ }\href {https://doi.org/10.1103/PhysRevB.110.L041113} {\bibfield  {journal} {\bibinfo  {journal} {Phys. Rev. B}\ }\textbf {\bibinfo {volume} {110}},\ \bibinfo {pages} {L041113} (\bibinfo {year} {2024})}\BibitemShut {NoStop}%
\bibitem [{\citenamefont {Chen}\ \emph {et~al.}(2022)\citenamefont {Chen}, \citenamefont {Chen}, \citenamefont {Gong}, \citenamefont {Sheng}, \citenamefont {Li},\ and\ \citenamefont {Weichselbaum}}]{PhysRevB.106.094420}%
  \BibitemOpen
  \bibfield  {author} {\bibinfo {author} {\bibfnamefont {B.-B.}\ \bibnamefont {Chen}}, \bibinfo {author} {\bibfnamefont {Z.}~\bibnamefont {Chen}}, \bibinfo {author} {\bibfnamefont {S.-S.}\ \bibnamefont {Gong}}, \bibinfo {author} {\bibfnamefont {D.~N.}\ \bibnamefont {Sheng}}, \bibinfo {author} {\bibfnamefont {W.}~\bibnamefont {Li}},\ and\ \bibinfo {author} {\bibfnamefont {A.}~\bibnamefont {Weichselbaum}},\ }\bibfield  {title} {\bibinfo {title} {Quantum spin liquid with emergent chiral order in the triangular-lattice hubbard model},\ }\href {https://doi.org/10.1103/PhysRevB.106.094420} {\bibfield  {journal} {\bibinfo  {journal} {Phys. Rev. B}\ }\textbf {\bibinfo {volume} {106}},\ \bibinfo {pages} {094420} (\bibinfo {year} {2022})}\BibitemShut {NoStop}%
\bibitem [{\citenamefont {Laubach}\ \emph {et~al.}(2015)\citenamefont {Laubach}, \citenamefont {Thomale}, \citenamefont {Platt}, \citenamefont {Hanke},\ and\ \citenamefont {Li}}]{PhysRevB.91.245125}%
  \BibitemOpen
  \bibfield  {author} {\bibinfo {author} {\bibfnamefont {M.}~\bibnamefont {Laubach}}, \bibinfo {author} {\bibfnamefont {R.}~\bibnamefont {Thomale}}, \bibinfo {author} {\bibfnamefont {C.}~\bibnamefont {Platt}}, \bibinfo {author} {\bibfnamefont {W.}~\bibnamefont {Hanke}},\ and\ \bibinfo {author} {\bibfnamefont {G.}~\bibnamefont {Li}},\ }\bibfield  {title} {\bibinfo {title} {Phase diagram of the hubbard model on the anisotropic triangular lattice},\ }\href {https://doi.org/10.1103/PhysRevB.91.245125} {\bibfield  {journal} {\bibinfo  {journal} {Phys. Rev. B}\ }\textbf {\bibinfo {volume} {91}},\ \bibinfo {pages} {245125} (\bibinfo {year} {2015})}\BibitemShut {NoStop}%
\bibitem [{\citenamefont {Hu}\ \emph {et~al.}(2019)\citenamefont {Hu}, \citenamefont {Zhu}, \citenamefont {Eggert},\ and\ \citenamefont {He}}]{diracj1j2}%
  \BibitemOpen
  \bibfield  {author} {\bibinfo {author} {\bibfnamefont {S.}~\bibnamefont {Hu}}, \bibinfo {author} {\bibfnamefont {W.}~\bibnamefont {Zhu}}, \bibinfo {author} {\bibfnamefont {S.}~\bibnamefont {Eggert}},\ and\ \bibinfo {author} {\bibfnamefont {Y.-C.}\ \bibnamefont {He}},\ }\bibfield  {title} {\bibinfo {title} {Dirac spin liquid on the spin-$1/2$ triangular heisenberg antiferromagnet},\ }\href {https://doi.org/10.1103/PhysRevLett.123.207203} {\bibfield  {journal} {\bibinfo  {journal} {Phys. Rev. Lett.}\ }\textbf {\bibinfo {volume} {123}},\ \bibinfo {pages} {207203} (\bibinfo {year} {2019})}\BibitemShut {NoStop}%
\bibitem [{\citenamefont {Gong}\ \emph {et~al.}(2019)\citenamefont {Gong}, \citenamefont {Zheng}, \citenamefont {Lee}, \citenamefont {Lu},\ and\ \citenamefont {Sheng}}]{chiralj1j2j3}%
  \BibitemOpen
  \bibfield  {author} {\bibinfo {author} {\bibfnamefont {S.-S.}\ \bibnamefont {Gong}}, \bibinfo {author} {\bibfnamefont {W.}~\bibnamefont {Zheng}}, \bibinfo {author} {\bibfnamefont {M.}~\bibnamefont {Lee}}, \bibinfo {author} {\bibfnamefont {Y.-M.}\ \bibnamefont {Lu}},\ and\ \bibinfo {author} {\bibfnamefont {D.~N.}\ \bibnamefont {Sheng}},\ }\bibfield  {title} {\bibinfo {title} {Chiral spin liquid with spinon fermi surfaces in the spin-$\frac{1}{2}$ triangular heisenberg model},\ }\href {https://doi.org/10.1103/PhysRevB.100.241111} {\bibfield  {journal} {\bibinfo  {journal} {Phys. Rev. B}\ }\textbf {\bibinfo {volume} {100}},\ \bibinfo {pages} {241111} (\bibinfo {year} {2019})}\BibitemShut {NoStop}%
\bibitem [{\citenamefont {Jiang}\ and\ \citenamefont {Jiang}(2020)}]{PhysRevLett.125.157002}%
  \BibitemOpen
  \bibfield  {author} {\bibinfo {author} {\bibfnamefont {Y.-F.}\ \bibnamefont {Jiang}}\ and\ \bibinfo {author} {\bibfnamefont {H.-C.}\ \bibnamefont {Jiang}},\ }\bibfield  {title} {\bibinfo {title} {Topological superconductivity in the doped chiral spin liquid on the triangular lattice},\ }\href {https://doi.org/10.1103/PhysRevLett.125.157002} {\bibfield  {journal} {\bibinfo  {journal} {Phys. Rev. Lett.}\ }\textbf {\bibinfo {volume} {125}},\ \bibinfo {pages} {157002} (\bibinfo {year} {2020})}\BibitemShut {NoStop}%
\bibitem [{\citenamefont {Peng}\ \emph {et~al.}(2021)\citenamefont {Peng}, \citenamefont {Jiang}, \citenamefont {Wang},\ and\ \citenamefont {Jiang}}]{Peng:2021aa}%
  \BibitemOpen
  \bibfield  {author} {\bibinfo {author} {\bibfnamefont {C.}~\bibnamefont {Peng}}, \bibinfo {author} {\bibfnamefont {Y.-F.}\ \bibnamefont {Jiang}}, \bibinfo {author} {\bibfnamefont {Y.}~\bibnamefont {Wang}},\ and\ \bibinfo {author} {\bibfnamefont {H.-C.}\ \bibnamefont {Jiang}},\ }\bibfield  {title} {\bibinfo {title} {Gapless spin liquid and pair density wave of the hubbard model on three-leg triangular cylinders},\ }\href {https://doi.org/10.1088/1367-2630/ac3a83} {\bibfield  {journal} {\bibinfo  {journal} {New Journal of Physics}\ }\textbf {\bibinfo {volume} {23}},\ \bibinfo {pages} {123004} (\bibinfo {year} {2021})}\BibitemShut {NoStop}%
\bibitem [{\citenamefont {Jiang}(2021)}]{dopetj}%
  \BibitemOpen
  \bibfield  {author} {\bibinfo {author} {\bibfnamefont {H.-C.}\ \bibnamefont {Jiang}},\ }\bibfield  {title} {\bibinfo {title} {Superconductivity in the doped quantum spin liquid on the triangular lattice},\ }\bibfield  {journal} {\bibinfo  {journal} {npj Quantum Materials}\ }\textbf {\bibinfo {volume} {6}},\ \href {https://doi.org/10.1038/s41535-021-00375-w} {10.1038/s41535-021-00375-w} (\bibinfo {year} {2021})\BibitemShut {NoStop}%
\bibitem [{\citenamefont {Zhu}\ \emph {et~al.}(2022)\citenamefont {Zhu}, \citenamefont {Sheng},\ and\ \citenamefont {Vishwanath}}]{dopehubbard}%
  \BibitemOpen
  \bibfield  {author} {\bibinfo {author} {\bibfnamefont {Z.}~\bibnamefont {Zhu}}, \bibinfo {author} {\bibfnamefont {D.~N.}\ \bibnamefont {Sheng}},\ and\ \bibinfo {author} {\bibfnamefont {A.}~\bibnamefont {Vishwanath}},\ }\bibfield  {title} {\bibinfo {title} {Doped mott insulators in the triangular-lattice hubbard model},\ }\bibfield  {journal} {\bibinfo  {journal} {Physical Review B}\ }\textbf {\bibinfo {volume} {105}},\ \href {https://doi.org/10.1103/physrevb.105.205110} {10.1103/physrevb.105.205110} (\bibinfo {year} {2022})\BibitemShut {NoStop}%
\bibitem [{\citenamefont {Xu}\ \emph {et~al.}(2024)\citenamefont {Xu}, \citenamefont {Zhu}, \citenamefont {Wu},\ and\ \citenamefont {Weng}}]{PhysRevB.109.L081116}%
  \BibitemOpen
  \bibfield  {author} {\bibinfo {author} {\bibfnamefont {J.-S.}\ \bibnamefont {Xu}}, \bibinfo {author} {\bibfnamefont {Z.}~\bibnamefont {Zhu}}, \bibinfo {author} {\bibfnamefont {K.}~\bibnamefont {Wu}},\ and\ \bibinfo {author} {\bibfnamefont {Z.-Y.}\ \bibnamefont {Weng}},\ }\bibfield  {title} {\bibinfo {title} {Hubbard model on a triangular lattice: The role of charge fluctuations},\ }\href {https://doi.org/10.1103/PhysRevB.109.L081116} {\bibfield  {journal} {\bibinfo  {journal} {Phys. Rev. B}\ }\textbf {\bibinfo {volume} {109}},\ \bibinfo {pages} {L081116} (\bibinfo {year} {2024})}\BibitemShut {NoStop}%
\bibitem [{\citenamefont {Gannot}\ \emph {et~al.}(2020)\citenamefont {Gannot}, \citenamefont {Jiang},\ and\ \citenamefont {Kivelson}}]{PhysRevB.102.115136}%
  \BibitemOpen
  \bibfield  {author} {\bibinfo {author} {\bibfnamefont {Y.}~\bibnamefont {Gannot}}, \bibinfo {author} {\bibfnamefont {Y.-F.}\ \bibnamefont {Jiang}},\ and\ \bibinfo {author} {\bibfnamefont {S.~A.}\ \bibnamefont {Kivelson}},\ }\bibfield  {title} {\bibinfo {title} {Hubbard ladders at small $u$ revisited},\ }\href {https://doi.org/10.1103/PhysRevB.102.115136} {\bibfield  {journal} {\bibinfo  {journal} {Phys. Rev. B}\ }\textbf {\bibinfo {volume} {102}},\ \bibinfo {pages} {115136} (\bibinfo {year} {2020})}\BibitemShut {NoStop}%
\bibitem [{\citenamefont {Shirakawa}\ \emph {et~al.}(2017)\citenamefont {Shirakawa}, \citenamefont {Tohyama}, \citenamefont {Kokalj}, \citenamefont {Sota},\ and\ \citenamefont {Yunoki}}]{PhysRevB.96.205130}%
  \BibitemOpen
  \bibfield  {author} {\bibinfo {author} {\bibfnamefont {T.}~\bibnamefont {Shirakawa}}, \bibinfo {author} {\bibfnamefont {T.}~\bibnamefont {Tohyama}}, \bibinfo {author} {\bibfnamefont {J.}~\bibnamefont {Kokalj}}, \bibinfo {author} {\bibfnamefont {S.}~\bibnamefont {Sota}},\ and\ \bibinfo {author} {\bibfnamefont {S.}~\bibnamefont {Yunoki}},\ }\bibfield  {title} {\bibinfo {title} {Ground-state phase diagram of the triangular lattice hubbard model by the density-matrix renormalization group method},\ }\href {https://doi.org/10.1103/PhysRevB.96.205130} {\bibfield  {journal} {\bibinfo  {journal} {Phys. Rev. B}\ }\textbf {\bibinfo {volume} {96}},\ \bibinfo {pages} {205130} (\bibinfo {year} {2017})}\BibitemShut {NoStop}%
\bibitem [{\citenamefont {Cookmeyer}\ \emph {et~al.}(2021)\citenamefont {Cookmeyer}, \citenamefont {Motruk},\ and\ \citenamefont {Moore}}]{PhysRevLett.127.087201}%
  \BibitemOpen
  \bibfield  {author} {\bibinfo {author} {\bibfnamefont {T.}~\bibnamefont {Cookmeyer}}, \bibinfo {author} {\bibfnamefont {J.}~\bibnamefont {Motruk}},\ and\ \bibinfo {author} {\bibfnamefont {J.~E.}\ \bibnamefont {Moore}},\ }\bibfield  {title} {\bibinfo {title} {Four-spin terms and the origin of the chiral spin liquid in mott insulators on the triangular lattice},\ }\href {https://doi.org/10.1103/PhysRevLett.127.087201} {\bibfield  {journal} {\bibinfo  {journal} {Phys. Rev. Lett.}\ }\textbf {\bibinfo {volume} {127}},\ \bibinfo {pages} {087201} (\bibinfo {year} {2021})}\BibitemShut {NoStop}%
\bibitem [{\citenamefont {Eugenio}\ \emph {et~al.}(2024)\citenamefont {Eugenio}, \citenamefont {Luo}, \citenamefont {Vishwanath},\ and\ \citenamefont {Volkov}}]{tunablehubbardmodel}%
  \BibitemOpen
  \bibfield  {author} {\bibinfo {author} {\bibfnamefont {P.~M.}\ \bibnamefont {Eugenio}}, \bibinfo {author} {\bibfnamefont {Z.-X.}\ \bibnamefont {Luo}}, \bibinfo {author} {\bibfnamefont {A.}~\bibnamefont {Vishwanath}},\ and\ \bibinfo {author} {\bibfnamefont {P.~A.}\ \bibnamefont {Volkov}},\ }\href {https://arxiv.org/abs/2406.02448} {\bibinfo {title} {Tunable $t-t'-u$ hubbard models in twisted square homobilayers}} (\bibinfo {year} {2024}),\ \Eprint {https://arxiv.org/abs/2406.02448} {arXiv:2406.02448 [cond-mat.str-el]} \BibitemShut {NoStop}%
\bibitem [{\citenamefont {C{\u a}lug{\u a}ru}\ \emph {et~al.}(2024)\citenamefont {C{\u a}lug{\u a}ru}, \citenamefont {Jiang}, \citenamefont {Hu}, \citenamefont {Pi}, \citenamefont {Yu}, \citenamefont {Vergniory}, \citenamefont {Shan}, \citenamefont {Felser}, \citenamefont {Schoop}, \citenamefont {Efetov}, \citenamefont {Mak},\ and\ \citenamefont {Bernevig}}]{newtwistedM}%
  \BibitemOpen
  \bibfield  {author} {\bibinfo {author} {\bibfnamefont {D.}~\bibnamefont {C{\u a}lug{\u a}ru}}, \bibinfo {author} {\bibfnamefont {Y.}~\bibnamefont {Jiang}}, \bibinfo {author} {\bibfnamefont {H.}~\bibnamefont {Hu}}, \bibinfo {author} {\bibfnamefont {H.}~\bibnamefont {Pi}}, \bibinfo {author} {\bibfnamefont {J.}~\bibnamefont {Yu}}, \bibinfo {author} {\bibfnamefont {M.~G.}\ \bibnamefont {Vergniory}}, \bibinfo {author} {\bibfnamefont {J.}~\bibnamefont {Shan}}, \bibinfo {author} {\bibfnamefont {C.}~\bibnamefont {Felser}}, \bibinfo {author} {\bibfnamefont {L.~M.}\ \bibnamefont {Schoop}}, \bibinfo {author} {\bibfnamefont {D.~K.}\ \bibnamefont {Efetov}}, \bibinfo {author} {\bibfnamefont {K.~F.}\ \bibnamefont {Mak}},\ and\ \bibinfo {author} {\bibfnamefont {B.~A.}\ \bibnamefont {Bernevig}},\ }\href {https://arxiv.org/abs/2411.18684} {\bibinfo {title} {A new moir\'e platform based on m-point twisting}} (\bibinfo {year} {2024}),\ \Eprint {https://arxiv.org/abs/2411.18684} {arXiv:2411.18684 [cond-mat.str-el]} \BibitemShut
  {NoStop}%
\bibitem [{\citenamefont {Kariyado}(2023)}]{twistedBC3}%
  \BibitemOpen
  \bibfield  {author} {\bibinfo {author} {\bibfnamefont {T.}~\bibnamefont {Kariyado}},\ }\bibfield  {title} {\bibinfo {title} {Twisted bilayer ${\text{bc}}_{3}$: Valley interlocked anisotropic flat bands},\ }\href {https://doi.org/10.1103/PhysRevB.107.085127} {\bibfield  {journal} {\bibinfo  {journal} {Phys. Rev. B}\ }\textbf {\bibinfo {volume} {107}},\ \bibinfo {pages} {085127} (\bibinfo {year} {2023})}\BibitemShut {NoStop}%
\bibitem [{\citenamefont {Bao}\ \emph {et~al.}(2024)\citenamefont {Bao}, \citenamefont {Wang}, \citenamefont {Liu},\ and\ \citenamefont {jing Wang}}]{jingwangmvalley}%
  \BibitemOpen
  \bibfield  {author} {\bibinfo {author} {\bibfnamefont {K.}~\bibnamefont {Bao}}, \bibinfo {author} {\bibfnamefont {H.}~\bibnamefont {Wang}}, \bibinfo {author} {\bibfnamefont {Z.}~\bibnamefont {Liu}},\ and\ \bibinfo {author} {\bibnamefont {jing Wang}},\ }\href {https://arxiv.org/abs/2412.19613} {\bibinfo {title} {Anisotropic band flattening in twisted bilayer of m-valley mxenes}} (\bibinfo {year} {2024}),\ \Eprint {https://arxiv.org/abs/2412.19613} {arXiv:2412.19613 [cond-mat.mes-hall]} \BibitemShut {NoStop}%
\bibitem [{\citenamefont {Lei}\ \emph {et~al.}(2024)\citenamefont {Lei}, \citenamefont {Mahon},\ and\ \citenamefont {MacDonald}}]{macdonaldMvalley}%
  \BibitemOpen
  \bibfield  {author} {\bibinfo {author} {\bibfnamefont {C.}~\bibnamefont {Lei}}, \bibinfo {author} {\bibfnamefont {P.~T.}\ \bibnamefont {Mahon}},\ and\ \bibinfo {author} {\bibfnamefont {A.~H.}\ \bibnamefont {MacDonald}},\ }\href {https://arxiv.org/abs/2411.18828} {\bibinfo {title} {Moir\'{e} band theory for m-valley twisted transition metal dichalcogenides}} (\bibinfo {year} {2024}),\ \Eprint {https://arxiv.org/abs/2411.18828} {arXiv:2411.18828 [cond-mat.mes-hall]} \BibitemShut {NoStop}%
\bibitem [{\citenamefont {Luo}\ \emph {et~al.}(2025)\citenamefont {Luo}, \citenamefont {Vishwanath},\ and\ \citenamefont {Kariyado}}]{luo2025singlebandsquarelatticehubbard}%
  \BibitemOpen
  \bibfield  {author} {\bibinfo {author} {\bibfnamefont {Z.-X.}\ \bibnamefont {Luo}}, \bibinfo {author} {\bibfnamefont {A.}~\bibnamefont {Vishwanath}},\ and\ \bibinfo {author} {\bibfnamefont {T.}~\bibnamefont {Kariyado}},\ }\href {https://arxiv.org/abs/2502.19483} {\bibinfo {title} {Single-band square lattice hubbard model from twisted bilayer c568}} (\bibinfo {year} {2025}),\ \Eprint {https://arxiv.org/abs/2502.19483} {arXiv:2502.19483 [cond-mat.str-el]} \BibitemShut {NoStop}%
\bibitem [{\citenamefont {Kalmeyer}\ and\ \citenamefont {Laughlin}(1987)}]{PhysRevLett.59.2095}%
  \BibitemOpen
  \bibfield  {author} {\bibinfo {author} {\bibfnamefont {V.}~\bibnamefont {Kalmeyer}}\ and\ \bibinfo {author} {\bibfnamefont {R.~B.}\ \bibnamefont {Laughlin}},\ }\bibfield  {title} {\bibinfo {title} {Equivalence of the resonating-valence-bond and fractional quantum hall states},\ }\href {https://doi.org/10.1103/PhysRevLett.59.2095} {\bibfield  {journal} {\bibinfo  {journal} {Phys. Rev. Lett.}\ }\textbf {\bibinfo {volume} {59}},\ \bibinfo {pages} {2095} (\bibinfo {year} {1987})}\BibitemShut {NoStop}%
\bibitem [{\citenamefont {Wen}\ \emph {et~al.}(1989)\citenamefont {Wen}, \citenamefont {Wilczek},\ and\ \citenamefont {Zee}}]{PhysRevB.39.11413}%
  \BibitemOpen
  \bibfield  {author} {\bibinfo {author} {\bibfnamefont {X.~G.}\ \bibnamefont {Wen}}, \bibinfo {author} {\bibfnamefont {F.}~\bibnamefont {Wilczek}},\ and\ \bibinfo {author} {\bibfnamefont {A.}~\bibnamefont {Zee}},\ }\bibfield  {title} {\bibinfo {title} {Chiral spin states and superconductivity},\ }\href {https://doi.org/10.1103/PhysRevB.39.11413} {\bibfield  {journal} {\bibinfo  {journal} {Phys. Rev. B}\ }\textbf {\bibinfo {volume} {39}},\ \bibinfo {pages} {11413} (\bibinfo {year} {1989})}\BibitemShut {NoStop}%
\bibitem [{\citenamefont {Schroeter}\ \emph {et~al.}(2007)\citenamefont {Schroeter}, \citenamefont {Kapit}, \citenamefont {Thomale},\ and\ \citenamefont {Greiter}}]{PhysRevLett.99.097202}%
  \BibitemOpen
  \bibfield  {author} {\bibinfo {author} {\bibfnamefont {D.~F.}\ \bibnamefont {Schroeter}}, \bibinfo {author} {\bibfnamefont {E.}~\bibnamefont {Kapit}}, \bibinfo {author} {\bibfnamefont {R.}~\bibnamefont {Thomale}},\ and\ \bibinfo {author} {\bibfnamefont {M.}~\bibnamefont {Greiter}},\ }\bibfield  {title} {\bibinfo {title} {Spin hamiltonian for which the chiral spin liquid is the exact ground state},\ }\href {https://doi.org/10.1103/PhysRevLett.99.097202} {\bibfield  {journal} {\bibinfo  {journal} {Phys. Rev. Lett.}\ }\textbf {\bibinfo {volume} {99}},\ \bibinfo {pages} {097202} (\bibinfo {year} {2007})}\BibitemShut {NoStop}%
\bibitem [{\citenamefont {Yao}\ and\ \citenamefont {Kivelson}(2007)}]{PhysRevLett.99.247203}%
  \BibitemOpen
  \bibfield  {author} {\bibinfo {author} {\bibfnamefont {H.}~\bibnamefont {Yao}}\ and\ \bibinfo {author} {\bibfnamefont {S.~A.}\ \bibnamefont {Kivelson}},\ }\bibfield  {title} {\bibinfo {title} {Exact chiral spin liquid with non-abelian anyons},\ }\href {https://doi.org/10.1103/PhysRevLett.99.247203} {\bibfield  {journal} {\bibinfo  {journal} {Phys. Rev. Lett.}\ }\textbf {\bibinfo {volume} {99}},\ \bibinfo {pages} {247203} (\bibinfo {year} {2007})}\BibitemShut {NoStop}%
\bibitem [{Note1()}]{Note1}%
  \BibitemOpen
  \bibinfo {note} {We exclude angles between $\protect \bm {a}_1$ and $\protect \bm {a}_2$ that result in square and isotropic triangular lattices.}\BibitemShut {Stop}%
\bibitem [{Note2()}]{Note2}%
  \BibitemOpen
  \bibinfo {note} {Layer group symmetry constrains the effective mass tensor to be diagonal along the principal axes.}\BibitemShut {Stop}%
\bibitem [{sup()}]{supp}%
  \BibitemOpen
  \href@noop {} {}\bibinfo {note} {See Supplemental Material.}\BibitemShut {Stop}%
\bibitem [{\citenamefont {Huang}\ \emph {et~al.}(2023)\citenamefont {Huang}, \citenamefont {Gong},\ and\ \citenamefont {Sheng}}]{PhysRevLett.130.136003}%
  \BibitemOpen
  \bibfield  {author} {\bibinfo {author} {\bibfnamefont {Y.}~\bibnamefont {Huang}}, \bibinfo {author} {\bibfnamefont {S.-S.}\ \bibnamefont {Gong}},\ and\ \bibinfo {author} {\bibfnamefont {D.~N.}\ \bibnamefont {Sheng}},\ }\bibfield  {title} {\bibinfo {title} {Quantum phase diagram and spontaneously emergent topological chiral superconductivity in doped triangular-lattice mott insulators},\ }\href {https://doi.org/10.1103/PhysRevLett.130.136003} {\bibfield  {journal} {\bibinfo  {journal} {Phys. Rev. Lett.}\ }\textbf {\bibinfo {volume} {130}},\ \bibinfo {pages} {136003} (\bibinfo {year} {2023})}\BibitemShut {NoStop}%
\bibitem [{\citenamefont {Huang}\ and\ \citenamefont {Sheng}(2022)}]{PhysRevX.12.031009}%
  \BibitemOpen
  \bibfield  {author} {\bibinfo {author} {\bibfnamefont {Y.}~\bibnamefont {Huang}}\ and\ \bibinfo {author} {\bibfnamefont {D.~N.}\ \bibnamefont {Sheng}},\ }\bibfield  {title} {\bibinfo {title} {Topological chiral and nematic superconductivity by doping mott insulators on triangular lattice},\ }\href {https://doi.org/10.1103/PhysRevX.12.031009} {\bibfield  {journal} {\bibinfo  {journal} {Phys. Rev. X}\ }\textbf {\bibinfo {volume} {12}},\ \bibinfo {pages} {031009} (\bibinfo {year} {2022})}\BibitemShut {NoStop}%
\bibitem [{\citenamefont {Venderley}\ and\ \citenamefont {Kim}(2019)}]{PhysRevB.100.060506}%
  \BibitemOpen
  \bibfield  {author} {\bibinfo {author} {\bibfnamefont {J.}~\bibnamefont {Venderley}}\ and\ \bibinfo {author} {\bibfnamefont {E.-A.}\ \bibnamefont {Kim}},\ }\bibfield  {title} {\bibinfo {title} {Density matrix renormalization group study of superconductivity in the triangular lattice hubbard model},\ }\href {https://doi.org/10.1103/PhysRevB.100.060506} {\bibfield  {journal} {\bibinfo  {journal} {Phys. Rev. B}\ }\textbf {\bibinfo {volume} {100}},\ \bibinfo {pages} {060506} (\bibinfo {year} {2019})}\BibitemShut {NoStop}%
\bibitem [{\citenamefont {Wu}\ \emph {et~al.}(2023)\citenamefont {Wu}, \citenamefont {Wu},\ and\ \citenamefont {Yao}}]{PhysRevLett.130.126001}%
  \BibitemOpen
  \bibfield  {author} {\bibinfo {author} {\bibfnamefont {Y.-M.}\ \bibnamefont {Wu}}, \bibinfo {author} {\bibfnamefont {Z.}~\bibnamefont {Wu}},\ and\ \bibinfo {author} {\bibfnamefont {H.}~\bibnamefont {Yao}},\ }\bibfield  {title} {\bibinfo {title} {Pair-density-wave and chiral superconductivity in twisted bilayer transition metal dichalcogenides},\ }\href {https://doi.org/10.1103/PhysRevLett.130.126001} {\bibfield  {journal} {\bibinfo  {journal} {Phys. Rev. Lett.}\ }\textbf {\bibinfo {volume} {130}},\ \bibinfo {pages} {126001} (\bibinfo {year} {2023})}\BibitemShut {NoStop}%
\bibitem [{\citenamefont {Jiang}\ \emph {et~al.}(2024)\citenamefont {Jiang}, \citenamefont {Petralanda}, \citenamefont {Skorupskii}, \citenamefont {Xu}, \citenamefont {Pi}, \citenamefont {C{\u a}lug{\u a}ru}, \citenamefont {Hu}, \citenamefont {Xie}, \citenamefont {Mustaf}, \citenamefont {H{\"o}hn}, \citenamefont {Haase}, \citenamefont {Vergniory}, \citenamefont {Claassen}, \citenamefont {Elcoro}, \citenamefont {Regnault}, \citenamefont {Shan}, \citenamefont {Mak}, \citenamefont {Efetov}, \citenamefont {Morosan}, \citenamefont {Kennes}, \citenamefont {Rubio}, \citenamefont {Xian}, \citenamefont {Felser}, \citenamefont {Schoop},\ and\ \citenamefont {Bernevig}}]{twistdatabase}%
  \BibitemOpen
  \bibfield  {author} {\bibinfo {author} {\bibfnamefont {Y.}~\bibnamefont {Jiang}}, \bibinfo {author} {\bibfnamefont {U.}~\bibnamefont {Petralanda}}, \bibinfo {author} {\bibfnamefont {G.}~\bibnamefont {Skorupskii}}, \bibinfo {author} {\bibfnamefont {Q.}~\bibnamefont {Xu}}, \bibinfo {author} {\bibfnamefont {H.}~\bibnamefont {Pi}}, \bibinfo {author} {\bibfnamefont {D.}~\bibnamefont {C{\u a}lug{\u a}ru}}, \bibinfo {author} {\bibfnamefont {H.}~\bibnamefont {Hu}}, \bibinfo {author} {\bibfnamefont {J.}~\bibnamefont {Xie}}, \bibinfo {author} {\bibfnamefont {R.~A.}\ \bibnamefont {Mustaf}}, \bibinfo {author} {\bibfnamefont {P.}~\bibnamefont {H{\"o}hn}}, \bibinfo {author} {\bibfnamefont {V.}~\bibnamefont {Haase}}, \bibinfo {author} {\bibfnamefont {M.~G.}\ \bibnamefont {Vergniory}}, \bibinfo {author} {\bibfnamefont {M.}~\bibnamefont {Claassen}}, \bibinfo {author} {\bibfnamefont {L.}~\bibnamefont {Elcoro}}, \bibinfo {author} {\bibfnamefont {N.}~\bibnamefont {Regnault}}, \bibinfo {author} {\bibfnamefont {J.}~\bibnamefont
  {Shan}}, \bibinfo {author} {\bibfnamefont {K.~F.}\ \bibnamefont {Mak}}, \bibinfo {author} {\bibfnamefont {D.~K.}\ \bibnamefont {Efetov}}, \bibinfo {author} {\bibfnamefont {E.}~\bibnamefont {Morosan}}, \bibinfo {author} {\bibfnamefont {D.~M.}\ \bibnamefont {Kennes}}, \bibinfo {author} {\bibfnamefont {A.}~\bibnamefont {Rubio}}, \bibinfo {author} {\bibfnamefont {L.}~\bibnamefont {Xian}}, \bibinfo {author} {\bibfnamefont {C.}~\bibnamefont {Felser}}, \bibinfo {author} {\bibfnamefont {L.~M.}\ \bibnamefont {Schoop}},\ and\ \bibinfo {author} {\bibfnamefont {B.~A.}\ \bibnamefont {Bernevig}},\ }\href {https://arxiv.org/abs/2411.09741} {\bibinfo {title} {2d theoretically twistable material database}} (\bibinfo {year} {2024}),\ \Eprint {https://arxiv.org/abs/2411.09741} {arXiv:2411.09741 [cond-mat.mtrl-sci]} \BibitemShut {NoStop}%
\bibitem [{\citenamefont {Pichler}\ \emph {et~al.}(2024)\citenamefont {Pichler}, \citenamefont {Kadow}, \citenamefont {Kuhlenkamp},\ and\ \citenamefont {Knap}}]{PhysRevB.110.045116}%
  \BibitemOpen
  \bibfield  {author} {\bibinfo {author} {\bibfnamefont {F.}~\bibnamefont {Pichler}}, \bibinfo {author} {\bibfnamefont {W.}~\bibnamefont {Kadow}}, \bibinfo {author} {\bibfnamefont {C.}~\bibnamefont {Kuhlenkamp}},\ and\ \bibinfo {author} {\bibfnamefont {M.}~\bibnamefont {Knap}},\ }\bibfield  {title} {\bibinfo {title} {Probing magnetism in moir\'e heterostructures with quantum twisting microscopes},\ }\href {https://doi.org/10.1103/PhysRevB.110.045116} {\bibfield  {journal} {\bibinfo  {journal} {Phys. Rev. B}\ }\textbf {\bibinfo {volume} {110}},\ \bibinfo {pages} {045116} (\bibinfo {year} {2024})}\BibitemShut {NoStop}%
\bibitem [{\citenamefont {Peri}\ \emph {et~al.}(2024)\citenamefont {Peri}, \citenamefont {Ilani}, \citenamefont {Lee},\ and\ \citenamefont {Refael}}]{PhysRevB.109.035127}%
  \BibitemOpen
  \bibfield  {author} {\bibinfo {author} {\bibfnamefont {V.}~\bibnamefont {Peri}}, \bibinfo {author} {\bibfnamefont {S.}~\bibnamefont {Ilani}}, \bibinfo {author} {\bibfnamefont {P.~A.}\ \bibnamefont {Lee}},\ and\ \bibinfo {author} {\bibfnamefont {G.}~\bibnamefont {Refael}},\ }\bibfield  {title} {\bibinfo {title} {Probing quantum spin liquids with a quantum twisting microscope},\ }\href {https://doi.org/10.1103/PhysRevB.109.035127} {\bibfield  {journal} {\bibinfo  {journal} {Phys. Rev. B}\ }\textbf {\bibinfo {volume} {109}},\ \bibinfo {pages} {035127} (\bibinfo {year} {2024})}\BibitemShut {NoStop}%
\bibitem [{\citenamefont {Khoo}\ \emph {et~al.}(2022)\citenamefont {Khoo}, \citenamefont {Pientka}, \citenamefont {Lee},\ and\ \citenamefont {Villadiego}}]{PhysRevB.106.115108}%
  \BibitemOpen
  \bibfield  {author} {\bibinfo {author} {\bibfnamefont {J.~Y.}\ \bibnamefont {Khoo}}, \bibinfo {author} {\bibfnamefont {F.}~\bibnamefont {Pientka}}, \bibinfo {author} {\bibfnamefont {P.~A.}\ \bibnamefont {Lee}},\ and\ \bibinfo {author} {\bibfnamefont {I.~S.}\ \bibnamefont {Villadiego}},\ }\bibfield  {title} {\bibinfo {title} {Probing the quantum noise of the spinon fermi surface with nv centers},\ }\href {https://doi.org/10.1103/PhysRevB.106.115108} {\bibfield  {journal} {\bibinfo  {journal} {Phys. Rev. B}\ }\textbf {\bibinfo {volume} {106}},\ \bibinfo {pages} {115108} (\bibinfo {year} {2022})}\BibitemShut {NoStop}%
\bibitem [{\citenamefont {Lee}\ and\ \citenamefont {Morampudi}(2023)}]{PhysRevB.107.195102}%
  \BibitemOpen
  \bibfield  {author} {\bibinfo {author} {\bibfnamefont {P.~A.}\ \bibnamefont {Lee}}\ and\ \bibinfo {author} {\bibfnamefont {S.}~\bibnamefont {Morampudi}},\ }\bibfield  {title} {\bibinfo {title} {Proposal to detect emergent gauge field and its meissner effect in spin liquids using nv centers},\ }\href {https://doi.org/10.1103/PhysRevB.107.195102} {\bibfield  {journal} {\bibinfo  {journal} {Phys. Rev. B}\ }\textbf {\bibinfo {volume} {107}},\ \bibinfo {pages} {195102} (\bibinfo {year} {2023})}\BibitemShut {NoStop}%
\bibitem [{\citenamefont {Chatterjee}\ \emph {et~al.}(2019)\citenamefont {Chatterjee}, \citenamefont {Rodriguez-Nieva},\ and\ \citenamefont {Demler}}]{PhysRevB.99.104425}%
  \BibitemOpen
  \bibfield  {author} {\bibinfo {author} {\bibfnamefont {S.}~\bibnamefont {Chatterjee}}, \bibinfo {author} {\bibfnamefont {J.~F.}\ \bibnamefont {Rodriguez-Nieva}},\ and\ \bibinfo {author} {\bibfnamefont {E.}~\bibnamefont {Demler}},\ }\bibfield  {title} {\bibinfo {title} {Diagnosing phases of magnetic insulators via noise magnetometry with spin qubits},\ }\href {https://doi.org/10.1103/PhysRevB.99.104425} {\bibfield  {journal} {\bibinfo  {journal} {Phys. Rev. B}\ }\textbf {\bibinfo {volume} {99}},\ \bibinfo {pages} {104425} (\bibinfo {year} {2019})}\BibitemShut {NoStop}%
\bibitem [{\citenamefont {Takei}\ and\ \citenamefont {Tserkovnyak}(2024)}]{PhysRevResearch.6.013043}%
  \BibitemOpen
  \bibfield  {author} {\bibinfo {author} {\bibfnamefont {S.}~\bibnamefont {Takei}}\ and\ \bibinfo {author} {\bibfnamefont {Y.}~\bibnamefont {Tserkovnyak}},\ }\bibfield  {title} {\bibinfo {title} {Detecting fractionalization in critical spin liquids using color centers},\ }\href {https://doi.org/10.1103/PhysRevResearch.6.013043} {\bibfield  {journal} {\bibinfo  {journal} {Phys. Rev. Res.}\ }\textbf {\bibinfo {volume} {6}},\ \bibinfo {pages} {013043} (\bibinfo {year} {2024})}\BibitemShut {NoStop}%
\bibitem [{\citenamefont {Banerjee}\ \emph {et~al.}(2023)\citenamefont {Banerjee}, \citenamefont {Zhu},\ and\ \citenamefont {Lin}}]{Banerjee:2023aa}%
  \BibitemOpen
  \bibfield  {author} {\bibinfo {author} {\bibfnamefont {S.}~\bibnamefont {Banerjee}}, \bibinfo {author} {\bibfnamefont {W.}~\bibnamefont {Zhu}},\ and\ \bibinfo {author} {\bibfnamefont {S.-Z.}\ \bibnamefont {Lin}},\ }\bibfield  {title} {\bibinfo {title} {Electromagnetic signatures of a chiral quantum spin liquid},\ }\href {https://doi.org/10.1038/s41535-023-00595-2} {\bibfield  {journal} {\bibinfo  {journal} {npj Quantum Materials}\ }\textbf {\bibinfo {volume} {8}},\ \bibinfo {pages} {63} (\bibinfo {year} {2023})}\BibitemShut {NoStop}%
\bibitem [{\citenamefont {Xie}\ \emph {et~al.}(2023)\citenamefont {Xie}, \citenamefont {Lantagne-Hurtubise}, \citenamefont {Young}, \citenamefont {Nadj-Perge},\ and\ \citenamefont {Alicea}}]{PhysRevLett.131.146601}%
  \BibitemOpen
  \bibfield  {author} {\bibinfo {author} {\bibfnamefont {Y.-M.}\ \bibnamefont {Xie}}, \bibinfo {author} {\bibfnamefont {E.}~\bibnamefont {Lantagne-Hurtubise}}, \bibinfo {author} {\bibfnamefont {A.~F.}\ \bibnamefont {Young}}, \bibinfo {author} {\bibfnamefont {S.}~\bibnamefont {Nadj-Perge}},\ and\ \bibinfo {author} {\bibfnamefont {J.}~\bibnamefont {Alicea}},\ }\bibfield  {title} {\bibinfo {title} {Gate-defined topological josephson junctions in bernal bilayer graphene},\ }\href {https://doi.org/10.1103/PhysRevLett.131.146601} {\bibfield  {journal} {\bibinfo  {journal} {Phys. Rev. Lett.}\ }\textbf {\bibinfo {volume} {131}},\ \bibinfo {pages} {146601} (\bibinfo {year} {2023})}\BibitemShut {NoStop}%
\bibitem [{\citenamefont {Thomson}\ \emph {et~al.}(2022)\citenamefont {Thomson}, \citenamefont {Sorensen}, \citenamefont {Nadj-Perge},\ and\ \citenamefont {Alicea}}]{PhysRevB.105.L081405}%
  \BibitemOpen
  \bibfield  {author} {\bibinfo {author} {\bibfnamefont {A.}~\bibnamefont {Thomson}}, \bibinfo {author} {\bibfnamefont {I.~M.}\ \bibnamefont {Sorensen}}, \bibinfo {author} {\bibfnamefont {S.}~\bibnamefont {Nadj-Perge}},\ and\ \bibinfo {author} {\bibfnamefont {J.}~\bibnamefont {Alicea}},\ }\bibfield  {title} {\bibinfo {title} {Gate-defined wires in twisted bilayer graphene: From electrical detection of intervalley coherence to internally engineered majorana modes},\ }\href {https://doi.org/10.1103/PhysRevB.105.L081405} {\bibfield  {journal} {\bibinfo  {journal} {Phys. Rev. B}\ }\textbf {\bibinfo {volume} {105}},\ \bibinfo {pages} {L081405} (\bibinfo {year} {2022})}\BibitemShut {NoStop}%
\bibitem [{\citenamefont {Luo}\ \emph {et~al.}(2024)\citenamefont {Luo}, \citenamefont {Qiu},\ and\ \citenamefont {Wu}}]{PhysRevB.109.L041103}%
  \BibitemOpen
  \bibfield  {author} {\bibinfo {author} {\bibfnamefont {X.-J.}\ \bibnamefont {Luo}}, \bibinfo {author} {\bibfnamefont {W.-X.}\ \bibnamefont {Qiu}},\ and\ \bibinfo {author} {\bibfnamefont {F.}~\bibnamefont {Wu}},\ }\bibfield  {title} {\bibinfo {title} {Majorana zero modes in twisted transition metal dichalcogenide homobilayers},\ }\href {https://doi.org/10.1103/PhysRevB.109.L041103} {\bibfield  {journal} {\bibinfo  {journal} {Phys. Rev. B}\ }\textbf {\bibinfo {volume} {109}},\ \bibinfo {pages} {L041103} (\bibinfo {year} {2024})}\BibitemShut {NoStop}%
\bibitem [{\citenamefont {Anderson}\ \emph {et~al.}(2023)\citenamefont {Anderson}, \citenamefont {Fan}, \citenamefont {Cai}, \citenamefont {Holtzmann}, \citenamefont {Taniguchi}, \citenamefont {Watanabe}, \citenamefont {Xiao}, \citenamefont {Yao},\ and\ \citenamefont {Xu}}]{Anderson:2023aa}%
  \BibitemOpen
  \bibfield  {author} {\bibinfo {author} {\bibfnamefont {E.}~\bibnamefont {Anderson}}, \bibinfo {author} {\bibfnamefont {F.-R.}\ \bibnamefont {Fan}}, \bibinfo {author} {\bibfnamefont {J.}~\bibnamefont {Cai}}, \bibinfo {author} {\bibfnamefont {W.}~\bibnamefont {Holtzmann}}, \bibinfo {author} {\bibfnamefont {T.}~\bibnamefont {Taniguchi}}, \bibinfo {author} {\bibfnamefont {K.}~\bibnamefont {Watanabe}}, \bibinfo {author} {\bibfnamefont {D.}~\bibnamefont {Xiao}}, \bibinfo {author} {\bibfnamefont {W.}~\bibnamefont {Yao}},\ and\ \bibinfo {author} {\bibfnamefont {X.}~\bibnamefont {Xu}},\ }\bibfield  {title} {\bibinfo {title} {Programming correlated magnetic states with gate-controlled moir{\'e}geometry},\ }\href {https://doi.org/10.1126/science.adg4268} {\bibfield  {journal} {\bibinfo  {journal} {Science}\ }\textbf {\bibinfo {volume} {381}},\ \bibinfo {pages} {325} (\bibinfo {year} {2023})}\BibitemShut {NoStop}%
\bibitem [{\citenamefont {Kivelson}\ \emph {et~al.}(1987)\citenamefont {Kivelson}, \citenamefont {Rokhsar},\ and\ \citenamefont {Sethna}}]{PhysRevB.35.8865}%
  \BibitemOpen
  \bibfield  {author} {\bibinfo {author} {\bibfnamefont {S.~A.}\ \bibnamefont {Kivelson}}, \bibinfo {author} {\bibfnamefont {D.~S.}\ \bibnamefont {Rokhsar}},\ and\ \bibinfo {author} {\bibfnamefont {J.~P.}\ \bibnamefont {Sethna}},\ }\bibfield  {title} {\bibinfo {title} {Topology of the resonating valence-bond state: Solitons and high-${T}_{c}$ superconductivity},\ }\href {https://doi.org/10.1103/PhysRevB.35.8865} {\bibfield  {journal} {\bibinfo  {journal} {Phys. Rev. B}\ }\textbf {\bibinfo {volume} {35}},\ \bibinfo {pages} {8865} (\bibinfo {year} {1987})}\BibitemShut {NoStop}%
\bibitem [{\citenamefont {Lee}\ \emph {et~al.}(2006)\citenamefont {Lee}, \citenamefont {Nagaosa},\ and\ \citenamefont {Wen}}]{RevModPhys.78.17}%
  \BibitemOpen
  \bibfield  {author} {\bibinfo {author} {\bibfnamefont {P.~A.}\ \bibnamefont {Lee}}, \bibinfo {author} {\bibfnamefont {N.}~\bibnamefont {Nagaosa}},\ and\ \bibinfo {author} {\bibfnamefont {X.-G.}\ \bibnamefont {Wen}},\ }\bibfield  {title} {\bibinfo {title} {Doping a mott insulator: Physics of high-temperature superconductivity},\ }\href {https://doi.org/10.1103/RevModPhys.78.17} {\bibfield  {journal} {\bibinfo  {journal} {Rev. Mod. Phys.}\ }\textbf {\bibinfo {volume} {78}},\ \bibinfo {pages} {17} (\bibinfo {year} {2006})}\BibitemShut {NoStop}%
\bibitem [{\citenamefont {Senthil}\ and\ \citenamefont {Fisher}(2001)}]{PhysRevB.64.214511}%
  \BibitemOpen
  \bibfield  {author} {\bibinfo {author} {\bibfnamefont {T.}~\bibnamefont {Senthil}}\ and\ \bibinfo {author} {\bibfnamefont {M.~P.~A.}\ \bibnamefont {Fisher}},\ }\bibfield  {title} {\bibinfo {title} {Detecting fractions of electrons in the high-${T}_{c}$ cuprates},\ }\href {https://doi.org/10.1103/PhysRevB.64.214511} {\bibfield  {journal} {\bibinfo  {journal} {Phys. Rev. B}\ }\textbf {\bibinfo {volume} {64}},\ \bibinfo {pages} {214511} (\bibinfo {year} {2001})}\BibitemShut {NoStop}%
\bibitem [{\citenamefont {Barkeshli}\ \emph {et~al.}(2014)\citenamefont {Barkeshli}, \citenamefont {Berg},\ and\ \citenamefont {Kivelson}}]{Barkeshli_2014}%
  \BibitemOpen
  \bibfield  {author} {\bibinfo {author} {\bibfnamefont {M.}~\bibnamefont {Barkeshli}}, \bibinfo {author} {\bibfnamefont {E.}~\bibnamefont {Berg}},\ and\ \bibinfo {author} {\bibfnamefont {S.}~\bibnamefont {Kivelson}},\ }\bibfield  {title} {\bibinfo {title} {Coherent transmutation of electrons into fractionalized anyons},\ }\href {https://doi.org/10.1126/science.1253251} {\bibfield  {journal} {\bibinfo  {journal} {Science}\ }\textbf {\bibinfo {volume} {346}},\ \bibinfo {pages} {722} (\bibinfo {year} {2014})}\BibitemShut {NoStop}%
\bibitem [{\citenamefont {Law}\ and\ \citenamefont {Lee}(2017)}]{Law:2017aa}%
  \BibitemOpen
  \bibfield  {author} {\bibinfo {author} {\bibfnamefont {K.~T.}\ \bibnamefont {Law}}\ and\ \bibinfo {author} {\bibfnamefont {P.~A.}\ \bibnamefont {Lee}},\ }\bibfield  {title} {\bibinfo {title} {1t-tas2 as a quantum spin liquid},\ }\href {https://doi.org/10.1073/pnas.1706769114} {\bibfield  {journal} {\bibinfo  {journal} {Proceedings of the National Academy of Sciences}\ }\textbf {\bibinfo {volume} {114}},\ \bibinfo {pages} {6996} (\bibinfo {year} {2017})}\BibitemShut {NoStop}%
\end{thebibliography}%

\begin{widetext}
\section{Supplemental Materials}
\setcounter{equation}{0}
\setcounter{figure}{0}
\setcounter{table}{0}
\makeatletter
\renewcommand{\theequation}{S\arabic{equation}}
\renewcommand{\thefigure}{S\arabic{figure}}
\renewcommand{\bibnumfmt}[1]{[S#1]}
\renewcommand{\citenumfont}[1]{S#1}
\subsection{A. The interlayer tunneling with two leading harmonics}
Due to spin $SU(2)$ symmetry, time-reversal symmetry $\Theta$ becomes effectively spinless. The low-energy states transform as,
\begin{equation}
\Theta|\bm{p},l,s\rangle=\nu|-\bm{p},l,s\rangle
\end{equation}
where $\nu$ is a phase factor and $s$ is the spin index. In momentum space, scattering amplitudes are specified by $T(\bm{Q},\bm{p})$. The interlayer tunneling is,
\begin{equation}
\sum_{\bm{Q},\bm{p}}c_{\bm{p},+}^\dagger c_{\bm{p}+\bm{Q},-}T(\bm{Q},\bm{p})+\text{H.c.}
\end{equation}
Within the two-center approximation \cite{macdonaldmodel}, the $\bm{p}$ dependence in $T(\bm{Q},\bm{p})$ is neglected, as the orbitals contributing to the monolayer band exhibit localization. A systematic small $\bm{p}$ expansion can capture the effects of extended Wannier functions \cite{newtwistedM}. We do not expect such high derivative terms to generate qualitatively new effects in the low-energy bands. Therefore, we consider only the leading constant term $T_{\bm{Q}}$. The time-reversal symmetry constrains that,
\begin{equation}
T_{\bm{Q}}=\bar{T}_{-\bm{Q}}
\end{equation}
In real space, interlayer tunneling is,
\begin{equation}
\int_{\bm{r}}c_{\bm{r},+}^\dagger c_{\bm{r},-}(2w_x\cos(\bm{Q}_x\cdot\bm{r}+\phi_x)+2w_y\cos(\bm{Q}_y\cdot\bm{r}+\phi_y)+\text{H.c.}
\end{equation}

\subsection{B. Characterization of anisotropic band structure}
\subsubsection{Parameterization of the Hamiltonian}
To theoretically explore the moiré Hamiltonian's general parameter space, we transform it into a convenient form by separating dimensionless parameters from dimensional quantities. $H_x$ and $H_y$ are,
\begin{equation}
H_{\mu}=\frac{\partial_\mu^2}{2m_\mu}+2w_\mu\cos(\bm{Q}_\mu\cdot\bm{r}),\mu=x,y
\end{equation}
Defining the dimensionless coordinates $\tilde{x}_\mu\equiv\bm{Q}_\mu\cdot\bm{r}/2$, we rewrite the Hamiltonian as, 
\begin{equation}
H_\mu=\frac{Q_\mu^2}{8m_\mu}\left(\tilde{\partial}_\mu^2+2\frac{8m_\mu w_\mu}{Q_\mu^2}\cos(2\tilde{x}_\mu)\right)
\end{equation}
This form leads us to reorganize the Hamiltonian's parameters as follows. We use the overbar notation for the geometric mean of quantities in the $x$ and $y$ directions, and $r$ for their ratios. The Hamiltonian can then be transformed into,
\begin{equation}
\begin{aligned}
H_x&=\frac{\bar{Q}^2}{8\bar{m}}\sqrt{r_K}\left(\tilde{\partial}_x^2+2\frac{8\bar{m}\bar{w}}{\bar{Q}^2}\sqrt{\frac{r_w}{r_K}}\cos(2\tilde{x})\right)\\
H_y&=\frac{\bar{Q}^2}{8\bar{m}}\frac{1}{\sqrt{r_K}}\left(\tilde{\partial}_y^2+2\frac{8\bar{m}\bar{w}}{\bar{Q}^2}\sqrt{\frac{r_K}{r_w}}\cos(2\tilde{y})\right)\\
\frac{\bar{Q}^2}{8\bar{m}}&=\frac{Q_xQ_y}{8\sqrt{m_xm_y}},\bar{w}=\sqrt{w_xw_y},r_K=\frac{Q_x^2/m_x}{Q_y^2/m_y},r_w=\frac{w_x}{w_y}
\end{aligned}
\end{equation}
$\bar{Q}^2/(8\bar{m})$ represents the overall moiré energy scale. In the small twist angle limit, the dimensionless quantity $8\bar{m}\bar{w}/\bar{Q}^2\thickapprox(\theta^*/\theta)^2$ governs the competition between kinetic energy and the moiré potential. The characteristic angle $\theta^*$ depends on the microscopic details. When the twist angle falls below $\theta^*$, the moiré potential dominates, resulting in a flat, isolated top band composed of orbitals localized around potential extrema. For illustrative purposes, we set the twist angle to a small value, $\theta=0.5\theta^*$, in the main text. The bandwidths of $H_{x}$ and $H_y$ can be determined analytically using Mathieu's equation, which is available in MATHEMATICA.

Because the interlayer tunneling energy scale is controlled by $\bar{w}$, we assume the tuning range of $D_z$ is approximately $|D_z|<O(10)\bar{w}$. At larger $D_z$, electrons polarize to one layer, weakening the moiré potential's effect. In this regime, the system behaves as a weakly interacting electron gas, lacking a strongly coupled flat band. The sign of $D_z$ is irrelevant, as the Hamiltonian at $-D_z$ is unitary equivalent to that at $D_z$ via a $\sigma_x$ transformation.

\subsubsection{The Bloch theorem from moiré translation symmetry}
The moiré Hamiltonian with a displacement field is,
\begin{equation}
H=
\begin{pmatrix}
h(-i\bm{\nabla})+\frac{D_z}{2}&T(\bm{r})\\
T(\bm{r})&h(-i\bm{\nabla})-\frac{D_z}{2}
\end{pmatrix}
\end{equation}
It displays translational symmetries,
\begin{equation}
\sigma_zT_{\bm{a}_{1m}}HT_{\bm{a}_{1m}}^{-1}\sigma_z=H,\sigma_zT_{\bm{a}_{2m}}HT_{\bm{a}_{2m}}^{-1}\sigma_z=H
\end{equation}
where $T_{\bm{a}_{1m}}|\bm{r},l\rangle\equiv|\bm{r}+\bm{a}_{1m},l\rangle,T_{\bm{a}_{2m}}|\bm{r},l\rangle\equiv|\bm{r}+\bm{a}_{2m},l\rangle$. Bloch's theorem states that the eigenstates of a translational invariant Hamiltonian are also eigenstates of the translational symmetry operators. Thus, the Bloch states $|\bm{k}\rangle$ of the top isolated band satisfy,
\begin{equation}
\sigma_z T_{\bm{a}_{1m}}|\bm{k}\rangle=e^{-i\bm{k}\cdot\bm{a}_{1m}}|\bm{k}\rangle,\sigma_zT_{\bm{a}_{2m}}|\bm{k}\rangle=e^{-i\bm{k}\cdot\bm{a}_{2m}}|\bm{k}\rangle
\end{equation}
The real-space wavefunction of the Bloch state is,
\begin{equation}
\begin{aligned}
\langle\bm{r}-\bm{a}_{1m},l|\bm{k}\rangle=le^{-i\bm{k}\cdot\bm{a}_{1m}}\langle\bm{r},l|\bm{k}\rangle,\langle\bm{r}-\bm{a}_{2m},l|\bm{k}\rangle=le^{-i\bm{k}\cdot\bm{a}_{2m}}\langle\bm{r},l|\bm{k}\rangle
\end{aligned}
\end{equation}
It follows that the plane wave basis for the Hamiltonian at momentum $\bm{k}$ is,
\begin{equation}
\begin{aligned}
&l=+,|\bm{k}+\bm{G}_m,+\rangle\\
&l=-,|\bm{k}+\bm{G}_m-\bm{Q}_x,-\rangle
\end{aligned}
\label{eq:momentumchoice}
\end{equation}

\subsubsection{The magnitude of hoppings vs displacement field}
For selected parameter $(r_K=0.7,r_w=3,\theta=0.5\theta^*)$, we study the change of hoppings as a function of displacement field $D_z$. The primary contribution to interchain hopping $t'$ comes from the real-space overlap of Wannier functions \cite{tunablehubbardmodel},
\begin{equation}
 t'=D_z\langle W_+|W_-\rangle 
\end{equation}
obtained by projecting the displacement field onto the top band. $W_{\pm}$ represents the Wannier function of Hamiltonian $H_{\pm}$. They are located at the two nearest-neighboring lattice points belonging to different chains. As shown in Fig.~\ref{fig:hoppingvsdz}, $t'$ linearly grows with $D_z$ over quite extended range ($0<D_z<4\bar{w}$). This implies that in this regime, perturbation around $D_z=0$ limit is a good approximation. This approximation can be further supported by the intrachain hopping $t$ remains almost unchanged under $D_z$ ($0<D_z<2\bar{w}$). This is consistent, as $t$ connects the orbitals with the same $\sigma_x$, restricting $D_z$ to appear as the second-order effect which necessarily involves interband process and is suppressed by the large band gap.
\begin{figure}
    \centering
    \includegraphics[width=0.5\linewidth]{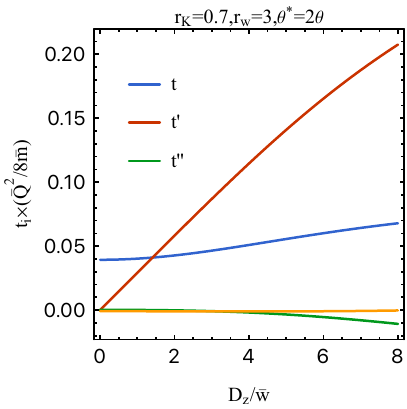}
    \caption{The displacement field effect on various hoppings}
    \label{fig:hoppingvsdz}
\end{figure}

\subsection{C. Other regimes of $r_K$}
We choose $r_K=1.5$ to illustrate a case where $r_K$ is near the isotropic bandwidth line without a displacement field. As shown in Fig.~\ref{fig:t'dominated}(a),
\begin{figure}
\includegraphics[width=1.0\linewidth]{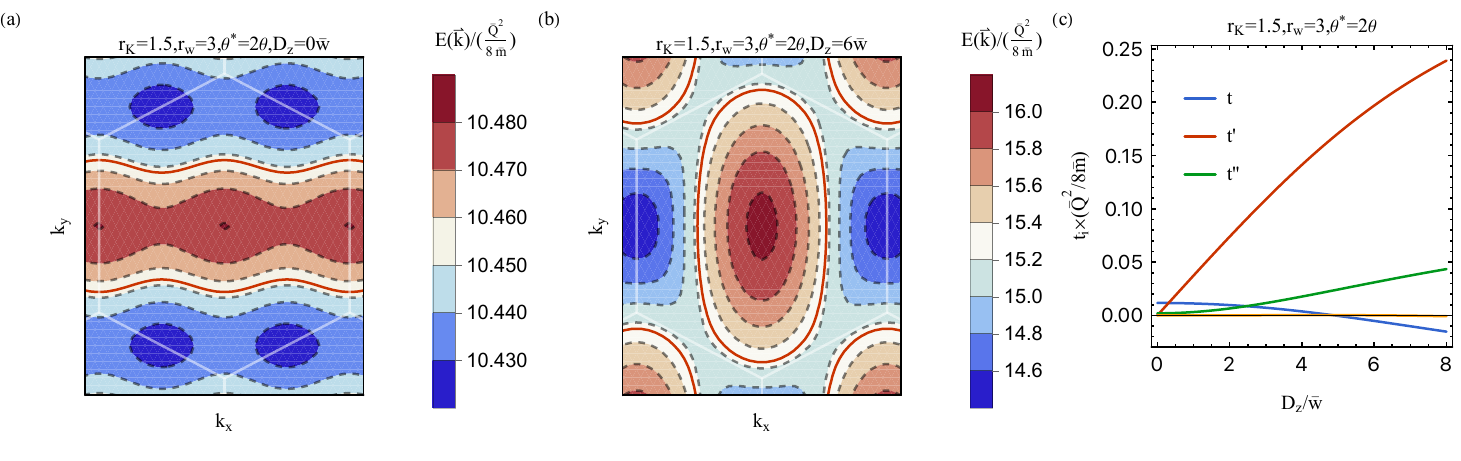}
\caption{\label{fig:t'dominated}Band structure for $r_K$ near the isotropic bandwidth line at zero displacement field. (a) Without the displacement field, band structure is anisotropic but not idealized one dimensional limit. (b) Without a displacement field, the band structure is anisotropic but not perfectly one-dimensional. (c) The band dispersion is well-fitted by intrachain hopping $t$, interchain nearest neighbor hopping $t'$ and interchain next-nearest neighbor hopping $t''$.}
\end{figure}
without a displacement field, the system can be described by quasi-one-dimensional chains for most fillings, except those near the band bottom top. With the displacement field applied (Fig.~\ref{fig:t'dominated}(c)), the nearest neighbor interchain hopping $t'$ rapidly increases and surpasses the intrachain hopping $t$ even at weak $D_z$. Simultaneously, the second nearest neighbor interchain hopping $t''$ also increases, while $t$ is suppressed. At large $D_z$, the dominant hopping is $t'$, but the subleading terms must include both $t''$ and $t$. The Fermi surface in this regime is shown in Fig.~\ref{fig:t'dominated}(b).

At large $r_K$, the system forms parallel $x$-directed chains. In this case, the interchain nearest neighbor distance is smaller than the intrachain nearest neighbor distance. In Fig.~\ref{xdirected},
\begin{figure}
\includegraphics[width=1.0\linewidth]{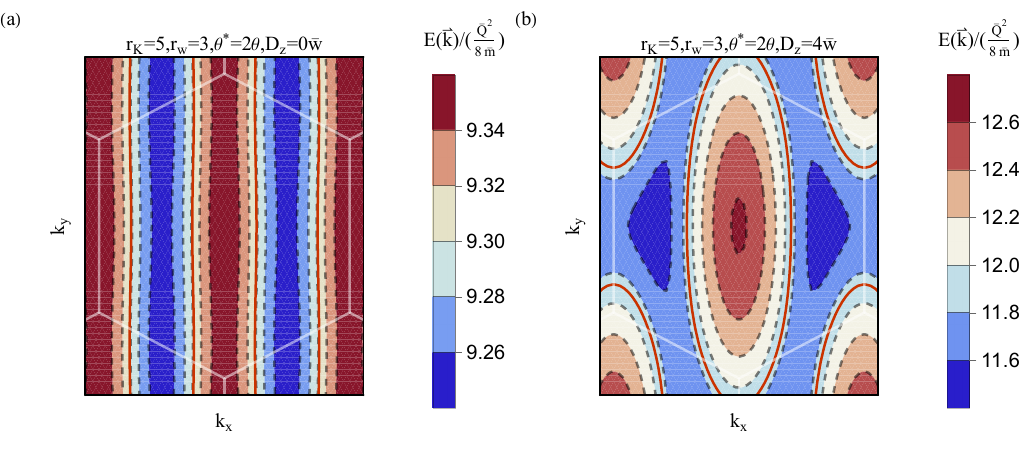}
\caption{\label{xdirected}Band structure of $x-$directed chains: (a) without and (b) with a displacement field.}
\end{figure}
we compare the Fermi surfaces of the top band as the displacement field changes. At $D_z=0$, there are four segments of one-dimensional Fermi surfaces. They satisfy the equation $E(\bm{k})=E(\bm{k}+\bm{Q}_x)$. This is a consequence of layer exchange symmetry \cite{newtwistedM,jingwangmvalley}. In real space, the layer exchange symmetry takes the form,
\begin{equation}
|\bm{r},l\rangle\to|\bm{r},-l\rangle
\end{equation}
Therefore, the symmetry action on momentum states of Eq.~\ref{eq:momentumchoice} is,
\begin{equation}
\begin{aligned}
|\bm{k}+\bm{G}-\frac{1-l}{2}\bm{Q}_x,l\rangle&\to|\bm{k}+\bm{G}-\frac{1-l}{2}\bm{Q}_x,-l\rangle\\
&=|\bm{k}+\bm{G}+l\bm{Q}_x-\frac{1+l}{2}\bm{Q}_x,-l\rangle
\end{aligned}
\end{equation}
This means that for layer $l$, we can shift the momentum as $\bm{G}\to\bm{G}-(l-1)\bm{Q}_x$. The Hamiltonian at momentum $\bm{k}$ is unitary equivalent to the Hamiltonian at momentum $\bm{k}+\bm{Q}_x$. They possess the same spectrum.

The one-dimensional Fermi surface corresponds to the  effective hopping $t''$ along the  $\pm(\bm{a}_{1m}+\bm{a}_{2m})$ direction. When the displacement field is applied, closed electron and hole pockets appear. The dispersions are well-fitted by the tight-binding model,
\begin{equation}
E(\bm{k})=2t''\cos(\bm{k}\cdot(\bm{a}_{1m}+\bm{a}_{2m}))+2t'\left(\cos(\bm{k}\cdot\bm{a}_{1m})+\cos(\bm{k}\cdot\bm{a}_{2m})\right)
\end{equation} 
Under the conditions depicted in Fig.~\ref{xdirected}(b), $t'/t''=0.93$.

\subsection{D. Real space structure of the band}
To gain further insight into the real-space structure of the top band, we plot the charge density distribution of the filled top band with and without the displacement field in Fig.~\ref{fig:bandrealspace}(a).
\begin{figure}
\includegraphics[width=1.0\linewidth]{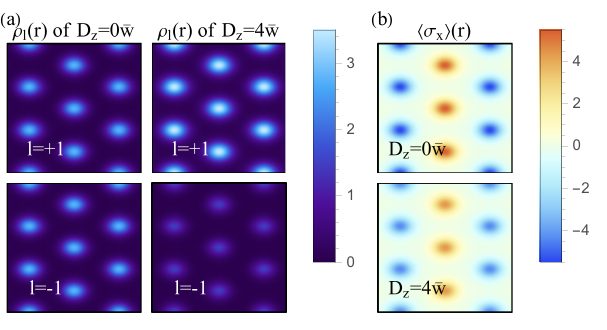}
\caption{\label{fig:bandrealspace}(a) Comparison of layer-resolved charge density from the top band at zero and nonzero displacement fields. (b) Comparison of the layer-pseudospin $\sigma_x$ density from the top band at zero and nonzero displacement fields.}
\end{figure}
The projection of continuum space operator $c_{\bm{r},l}$ onto the top band is,
\begin{equation}
\begin{aligned}
c_{\bm{r},l}&=\frac{1}{\sqrt{N_k}}\sum_{\bm{k}\in\text{1mBZ}}\psi_{\bm{k}}(\bm{r},l)c_{\bm{k}}\\
&=\frac{1}{\sqrt{N_k\Omega_{\text{mUC}}}}\sum_{\bm{k}\in\text{1mBZ},\bm{G}}e^{i(\bm{k}+\bm{G}-(1-l)/2\bm{Q}_x)\cdot\bm{r}}u_{\bm{k}}(\bm{G},l)c_{\bm{k}}
\end{aligned}
\end{equation}
where $N_k$ is the number of moiré unit cell in the system, $\Omega_{\text{mUC}}$ is the moiré unit cell size and $c_{\bm{k}}$ is annihilation operator for a state with crystal momentum $\bm{k}$ in the top band. The layer-resolved charge density is,
\begin{equation}
\begin{aligned}
&\rho_l(\bm{r})=\langle c_{\bm{r},l}^\dagger c_{\bm{r},l}\rangle\\
=&\frac{1}{N_k\Omega_{\text{mUC}}}\sum_{\bm{k}\in\text{1mBZ},\bm{G},\bm{G}'}e^{i(\bm{G}-\bm{G}')\cdot\bm{r}}\bar{u}_{\bm{k}}(\bm{G}',l)u_{\bm{k}}(\bm{G},l)
\end{aligned}
\end{equation}
With zero displacement field, the charge is evenly distributed over two layers and concentrated at the extrema of $T(\bm{r})$. With a positive displacement field, the charge density is pushed into the top layer. This aligns with the intuition that low-potential regions are occupied by lower bands, forcing top-band electrons into high-potential regions. The top band exhibits a nontrivial layer pseudospin $\sigma_x$ texture, characterizing its bonding and antibonding content (Fig.~\ref{fig:bandrealspace}(b)). 
\begin{equation}
\begin{aligned}
&\langle\sigma_x\rangle(\bm{r})=\sum_l\langle c_{\bm{r},-l}^\dagger c_{\bm{r},l}\rangle\\
=&\frac{1}{N_k\Omega_{\text{mUC}}}\sum_{\bm{k}\in\text{1mBZ},\bm{G},\bm{G}',l}e^{i(\bm{G}-\bm{G}'+\bm{Q}_xl)\cdot\bm{r}}\bar{u}_{\bm{k}}(\bm{G}',-l)u_{\bm{k}}(\bm{G},l)
\end{aligned}
\end{equation}
The momentum $\bm{G}+\bm{Q}_x$ indicates a pattern periodicity of $\bm{a}_{2m}\pm\bm{a}_{1m}$ and a sign change in $\sigma_x$ upon translation by $\bm{a}_{1m}$, i.e., $\sigma_x(\bm{r}+\bm{a}_{1m})=-\sigma_x(\bm{r})$. Without a displacement field, the pseudospin density peaks around the Wannier centers due to exact layer exchange symmetry. The displacement field adiabatically weakens the pseudospin density.

\subsection{E. Wannier function construction}
Motivated by the layer pseudospin structure of the top band, we aim to construct the localized Wannier function. We fix the momentum space gauge by requiring the Wannier function at lattice point $\bm{R}=\bm{0}$ to have positive weight on the layer exchange symmetric state at $\bm{r}=\bm{0}$,
\begin{equation}
(\langle\bm{r}=0,+|+\langle\bm{r}=0,-|)|\bm{R}=0\rangle>0
\end{equation}
In the momentum space, this requirement becomes,
\begin{equation}
\begin{aligned}
&(\langle\bm{r}=0,+|+\langle\bm{r}=0,-|)|\bm{R}=0\rangle\\
=&\frac{1}{\sqrt{N_k\Omega}}\sum_{\bm{k},\bm{G},l}u_{\bm{k}}(\bm{G},l)>0
\end{aligned}
\end{equation}
To satisfy this requirement, the gauge choice for $|\bm{k}\rangle$ at each momentum $\bm{k}$ is defined as,
\begin{equation}
\sum_{\bm{G},l}u_{\bm{k}}(\bm{G},l)>0
\end{equation}

\end{widetext}

\end{document}